**Speeds of sound for ($CH_4$ + He) mixtures from $p$ = (0.5 to 20) MPa at $T$ = (273.16 to 375) K**

Daniel Lozano-Martín[a], Andres Rojo[b], M. Carmen Martín[a], David Vega-Maza[c], José Juan Segovia[a]

[a] BioEcoUVa, Research Institute on Bioeconomy, TERMOCAL Research Group, University of Valladolid, Escuela de Ingenierias Industriales, Paseo del Cauce 59, 47011 Valladolid, Spain

[b] CEM (Centro Español de Metrología), Reference Gas Laboratory, Alfar 2, 28760 Tres Cantos, Madrid, Spain.

[c] School of Engineering, University of Aberdeen, Aberdeen, AB24 3FX, United Kingdom.

## Abstract

This work aims to provide accurate and wide-ranging experimental new speed of sound data $w(p,T)$ of two binary ($CH_4$ + He) mixtures at a nominal helium content of 5 % and 10 % at pressures $p$ = (0.5 up to 20) MPa and temperatures $T$ = (273.16, 300, 325, 350 and 375) K. For this purpose, the most accurate technique for determining speed of sound in gas phase has been used: the spherical acoustic resonator. Speed of sound is determined with an overall relative expanded ($k$ = 2) uncertainty of 230 parts in $10^6$ and compared to reference models for multicomponent natural gas-like mixtures: AGA8-DC92 and GERG-2008 equations of state. Relative deviations of experimental data from model estimations are outside the experimental uncertainty limit, although all points are mostly within the AGA uncertainty of 0.2 % and GERG uncertainty of 0.5 % and worsen as the helium content increases. Absolute average deviations are better than 0.45 % for GERG and below 0.14 % for AGA models in (0.95 $CH_4$ + 0.05 He) mixture and below 0.83 % for GERG and within 0.22 % for AGA equations in (0.90 $CH_4$ + 0.10 He) mixture.

**Keywords**

Speed of sound; acoustic resonance; methane; helium; heat capacities as perfect gas; virial coefficients.

| *Nomenclature* | | *Greek symbols* | |
|---|---|---|---|
| $a$ | Inner radius of the cavity, m | $\alpha$ | Reduced Helmholtz free energy<br>Thermal expansion coefficient, $K^{-1}$ |
| $A_i$ | Coefficients of acoustic virial equation | $\beta_a$ | 2nd acoustic virial coefficient, $m^3 \cdot mol^{-1}$ |
| $b$ | Outer radius of the cavity, m | $\Delta$ | Frequency perturbation, Hz |
| $B(T)$ | Second virial coefficient, $cm^3 \cdot mol^{-1}$ | $\gamma$ | Adiabatic coefficient |
| $C_p$ | Isobaric heat capacity, $J \cdot kg^{-1} \cdot K^{-1}$ | $\gamma_a$ | 3rd acoustic virial coefficient, $m^6 \cdot mol^{-2}$ |
| $C_{p,w}$ | Isobaric heat capacity of the wall material, $J \cdot kg^{-1} \cdot K^{-1}$ | $\gamma_{eff}$ | Effective adiabatic coefficient |
| $C_V$ | Isochoric heat capacity, $J \cdot kg^{-1} \cdot K^{-1}$ | $\delta$ | Reduced density |
| $E$ | Young's modulus, Pa | $\eta$ | Shear viscosity, Pa·s |
| $f$ | Resonance frequency, Hz | $\kappa$ | Thermal conductivity, $W \cdot m^{-1} \cdot K^{-1}$ |
| $g$ | Resonance halfwidth, Hz<br>Gravitational acceleration, $m \cdot s^{-2}$ | $\kappa_w$ | Thermal conductivity of the wall material, $W \cdot m^{-1} \cdot K^{-1}$ |
| $h$ | Thermal accommodation coefficient | $\kappa_T$ | Isothermal compressibility, $Pa^{-1}$ |
| $h_P$ | Planck Constant, J·s | $\nu_{0n}$ | Acoustic radial mode eigenvalue |
| $k$ | Coverage factor | $\nu_i$ | Molecular vibrational frequency, Hz |
| $k_B$ | Boltzmann Constant, $J \cdot K^{-1}$ | $\rho$ | Density, $kg \cdot m^{-3}$ |
| $m$ | Mass, kg | $\rho_n$ | Molar density, $mol \cdot m^{-3}$ |
| L | Duct length, m | $\rho_w$ | Density of the wall material, $kg \cdot m^{-3}$ |
| M | Molar Mass, kg/mol | $\sigma$ | Poisson's ratio |
| N | Number of components of a mixture | $\tau$ | Reduced temperature |
| $p$ | Pressure, MPa | $\tau_{vib}$ | Vibrational relaxation constant, s |
| $r_0$ | Duct radius, m | *Subscripts* | |
| $r_{tr}$ | Radius of the transducer, m | 0 | Reference state |
| $R$ | Molar gas constant, $J \cdot mol^{-1} \cdot K^{-1}$ | 0n | Acoustic radial mode index |
| $s$ | Standard deviation | 1 | Component 1 of a binary mixture |
| $T$ | Temperature, K | 2 | Component 2 of a binary mixture |
| $u$ | Standard uncertainty | AGA | Calculated from AGA equation of state |
| $U$ | Expanded uncertainty | c | Critical parameter |
| $V$ | Volume, $m^3$ | EoS | Calculated from an equation of state |
| $V_h$ | Volume of the holes drilled in the transducer backplate, $m^3$ | exp | Experimental data |
| $w$ | Speed of sound, $m \cdot s^{-1}$ | GERG | Calculated from GERG equation of state |
| $w_w$ | Speed of sound in the wall material, $m \cdot s^{-1}$ | r | Relative |
| $Z$ | Compressibility factor | th | Thermal boundary layer |
| *Abbreviations* | | sh | Shell |
| BAM | Federal Institute for Materials Research and Testing | *Superscripts* | |
| CEM | National Metrology Institute of Spain | 0, *pg* | Ideal gas behavior |
| GUM | Guide to the Expression of Uncertainty in Measurement | r | Residual behavior |

## 1. Introduction.

Helium is currently the subject of much attention. This noble gas is widely used in medical, scientific, aerospace, and electronic applications such that demand for it is expected to exceed supply in the coming years [1] with prices suffering substantial increases. This finite resource is obtained from natural gas reservoirs, although only when the concentration is above 0.2 % does recovery from natural gas by cryogenic and distillation methods or liquefaction plants prove economically viable [2]. Accurate thermodynamic models are required to design the recovery of helium from natural gas and for the transportation, storage, and liquefaction stages. Reference thermodynamic models are the AGA8-DC92 [3] and the GERG-2008 [4] equations of state (EoS). Both models are based on a multi-fluid expression of the Helmholtz free energy as a function of density $\rho$, temperature $T$, and composition $x$, from which all other thermodynamic properties can be estimated. The Helmholtz free energy is divided into two contributions: the ideal gas and the residual (real gas) part. The AGA8 model represents the ideal part of the Helmholtz free energy from the isobaric heat capacity in the ideal gas state and describes the residual part of the Helmholtz free energy from the AGA-8 mixture model with binary interaction parameters fitted from properly selected experimental data. The residual part is specifically written in the reduced dimensionless temperature $\tau$ and reduced dimensionless density $\delta$: $\tau = L / T$ with $L$ = 1 K, and $\delta = K^3 \rho$ where $K$ is a mixture size parameter. The GERG-2008 model represents the ideal part of the Helmholtz free energy in the same way as the AGA-8 EoS. The difference is that the residual part of the Helmholtz free energy has mixture parameters in both the reduced temperature and reduced density. In addition, for components with accurate binary mixture data, specific or generalized departure functions are added to the residual part to increase the model accuracy of the GERG EoS. The departure functions are less significant terms that depend on mixture composition.

Thus, accurate and extensive experimental data are required to fit the parameters of the correlation equations that describe the binary interaction between the substances of a real multicomponent mixture. In the case of helium, scarce data are available; the GERG-2008 EoS only uses $CH_4$ - He vapour-liquid equilibrium (VLE) data to model interaction with methane, the most important natural gas component. Moreover, GERG-2008 EoS states that it is worth developing a generalized departure function of binary mixtures containing helium in order to increase the accuracy of the model when suitable accurate data become available. This encouraged us to provide comprehensive and accurate speed of sound *w* data for two methane and helium binary mixtures with the amount of substance fractions (0.95 $CH_4$ + 0.05 He) and (0.90 $CH_4$ + 0.10 He) in this work.

Speed of sound measurements are performed using an acoustic spherical resonator at pressures from 0.5 MPa up to 20 MPa, and temperatures in the standard range for natural gas applications at (273.16, 300, 325, 350 and 375) K. This is the most accurate method for measuring speed of sound in gases, as can be deduced from its key role in the recent re-determination of the Boltzmann constant [5-10]. Results are compared to both AGA-8 and GERG-2008 EoS. AGA-8 EoS states general uncertainties with a 95 % confidence interval ($k$ = 2) for speed of sound, ranging up to 0.2 % at pressures below 5 MPa, 0.8 % between (5 to 10) MPa, and up to 2.0 % for higher pressures. GERG-2008 EoS estimates general uncertainties for the speed of sound of binary mixture without a departure function of 1 % for temperatures below 285 K and 0.5 % for temperatures up to 420 K in the pressure range of this work. No speed of sound data were found in the literature for binary mixtures of methane with helium, although high quality speed of sound measurements have been performed for pure methane [11-14] and pure helium [15]. The setup of this research has also been used successfully in previous works [16-17] and [18] for other binary and natural gas-like mixtures.

## 2. Experimental setup.

### 2.1. Acoustic resonator.

The acoustic spherical cavity used in this work is shown in figure 1. It was designed and made at Imperial College London in austenitic stainless-steel 321 grade by electron beam welding of two aligned hemispheres [19-20]. The sphere has an internal nominal radius $a$ = 40 mm and an external nominal radius $b$ = 52.5 mm. The internal radius calibration procedure as a function of pressure and temperature is described in section 4.1. The hemispheres were machined to the best possible tolerance and polished from a stainless-steel bar stock using a turret lathe. The welding penetrates until half of the wall thickness. However, the equatorial junction gap and other geometry imperfections such as spheroidal distortions, misalignment and unequal radii of the two hemispheres are second order perturbations that have been shown to have a negligible effect on the radial acoustic modes within the accuracy of this research [21-22]. The sealing gaskets of the transducer ports and upper and lower gas ducts are made of Kalrez perfluoroelastomer.

The acoustic wave is produced and detected by two non-commercial and equal acoustic transducers of solid dielectric capacitance type [23]. They are located flush with the internal surface of the acoustic cavity at 45º of the north pole with a 90º separation between them. This position was chosen at the design stage to reduce overlapping of the fundamental acoustic resonance mode (0,2) and the close mode (3,1). These are devices of wide frequency bandwidth and high acoustic impedance to the fluid, consisting of a polyamide dielectric diaphragm of 12 μm thickness and 3 mm diameter, gold plated on the external side by a 50 nm thick layer. The electrical circuit is closed by a steel backplate drilled to increase acoustic sensitivity. The theoretical mechanical frequencies of this assembly should be around 40 kHz, well above the acoustic resonances of this work.

The source transducer is driven by an alternate signal without offset produced by a wave synthesizer (model 3225B, HP), and sound is produced at twice the selected frequency, avoiding the undesirable crosstalk effect. The 40 peak-to-peak voltage that feeds the source transducer is increased to 180 V after passing an impedance adapter.

The detector transducer is fed with a bias voltage of 80 V and operated at constant charge by connection to a high input impedance and unity gain amplifier. The input connection to the detector transducer is made using triaxial cables in active guard configuration to remove the capacitance effect of the connection cables, which is several times higher than the 100 pF capacitance of the transducer.

The output signal of the amplifier is measured as in-phase and quadrature signals by a Lock-In detector (model SR850 DSP, Stanford Research Systems), which is referenced to the second harmonic of the wave generator. The complex signal $z(f) = A(\cos\varphi + i\,\mathrm{sen}\,\varphi)$ is scanned at 11 equally spaced drive frequencies $f$ in a ramp up and down centred around the theoretical resonance frequency and fitted to a Lorentzian shape function with a linear background level:

$$z(f) = \frac{A^*}{\left(F^2 - f^2\right)} + B^* + C^* f \qquad (1)$$

where $A$ is the measured signal amplitude, $\varphi$ is its phase, $A^*$ is a complex fitted parameter, $F = f_0 + ig$, $f_0$ is the resonance frequency, $g$ is the resonance halfwidth, and $B^*$ and $C^*$ are complex parameters. The fit is implemented in Agilent VEE 7.0 software following the nonlinear regression algorithm of Mehl [24], and Ewing and Trusler [25] with $C^* = 0$ and $C^* \neq 0$, respectively. The results with least regression error are chosen as the experimental resonance frequency and halfwidth.

### 2.2. Temperature control.

Cavity temperature is measured by two 25.5 Ω SPRTs (standard platinum resistance thermometers) (model 162D, Rosemount) located in mounting blocks on the north and south hemispheres, respectively, and measured by an AC bridge (model ASL F18, Automatic Systems Laboratories) referenced to an external 100 Ω resistance (model 5685A, Tinsley). Temperature stability is achieved by introducing the resonance cavity into a thermostat consisting of an ethanol Dewar cooled by a thermal bath (model FP89, Julabo), and an external shell, an internal shell, and a copper block from which the cavity hangs (figure 1). A vacuum is created inside the external and internal shell, avoiding heat transfer by convection using a turbomolecular (model SL300, Oerlikon) and a rotatory (model Trivac B8B, Leybold) pump. Additionally, several aluminium foils surrounding the internal shell limit heat transfer by radiation. Three proportional + integral + derivative (PID) controllers comprising three heating resistors and Pt-25 SPRTs (model S1509, Minco) are located on the copper block at the side of the internal vessel and at the base of the internal vessel, respectively. Resonance cavity temperature is thus only set by heat conduction through the copper block, and the thermal gradient between hemispheres is reduced to a few mK.

### 2.3. Pressure control.

The spherical resonator also works as the pressure-tight vessel. Pressure is gauged with two piezoelectric quartz transducers, (model 43KR-101, Digiquartz) for pressures above 2 MPa and a (model 2003A-101, Digiquartz) for pressures below this point. Both are located at the top of the gas inlet tube and in direct contact with the gas sample. The temperatures given by two thermocouples spaced across the length of the inlet tube and the temperature given by the pressure transducer itself are used to correct the cavity pressure by the hydrostatic column. Pressure is achieved after several loads from the gas bottle using a hand operated piston pump. Between each measurement point, pressure is reduced by venting the gas sample to ambient.

### 2.4. Mixture preparation.

The methane + helium-4 binary mixtures used in this work were synthesized at the BAM (Bundesanstalt für Materialforschung und -prüfung) Federal Institute for Materials Research and Testing by the gravimetric method in accordance with EN ISO 6142:2006 [26] and validated by gas chromatography. The composition of the mixtures is given in Table 1. The critical points computed from RefProp [27] are: $T_C$ = 194.24 K, $p_C$ = 6.4381 MPa, and $\rho_C$ = 160.19 kg m$^{-3}$ for the (0.95 $CH_4$ + 0.05 He) mixture and $T_C$ = 196.25 K, $p_C$ = 8.2372 MPa, and $\rho_C$ = 163.12 kg m$^{-3}$ for the (0.90 $CH_4$ + 0.10 He) mixture. The pure methane and helium-4 used to prepare the mixture were supplied by Linde AG (Germany) with a specified purity of over 0.999995 mol/mol and 0.999999 mol/mol, respectively. Details of the preparation method and its validation are described elsewhere [28]. The source of the compounds used is reported in Table 2, and the final purity and analytical method used for sample characterization is given in Table 3. This table reports the results of the check of the molar composition of the mixtures by a gas chromatography (GC) analysis. In any case the reported data should be attributed to the composition of Table 1. Mixtures were homogenized again by rolling prior to performing the measurements conducted in this research.

## 3. Data analysis.

### 3.1. Acoustic model.

Applying the boundary radial conditions to the solution of the homogeneous Helmholtz wave equation with the assumptions of zero surface acoustic admittance and perfect geometry leads to the expression that relates the experimental acoustic resonance frequency $f_{0n}$ to speed of sound $w(p,T)$ in the fluid [21]:

$$w(p,T) = \frac{2\pi a}{\nu_{0n}} f_{0n} \tag{2}$$

where $\nu_{0n}$ is the zero of the spherical Bessel first derivative of the $n$-th mode of order $l = 0$, and $a$ is the internal radius of the cavity at each pressure and temperature. In this ideal case, the only contribution to the experimental halfwidth $g$ is due to the classical viscothermal dissipation of acoustic energy in the fluid bulk $g_{cl}$:

$$g_{cl} = f^3 \frac{\pi^2}{w^2}\left[\frac{4}{3}\delta_s^2 + (\gamma - 1)\delta_{th}^2\right] \tag{3}$$

where $\gamma$ is the adiabatic coefficient, and the thickness of the viscous boundary layer $\delta_S$ and the thickness of the thermal boundary layer in the fluid $\delta_{th}$ are:

$$\delta_S = \left[\eta/(\rho\pi f)\right]^{1/2} \tag{4}$$

$$\delta_{th} = \left[\kappa/(\pi\rho C_p f)\right]^{1/2} \tag{5}$$

with $\eta$ being the shear viscosity, $\rho$ the density, $\kappa$ the thermal conductivity, and $C_p$ the isobaric heat capacity of the fluid. First order perturbation theory is applied to evaluate the frequency shifts $\Delta f$ that must be subtracted to $f_{0n}$ to account for the different effects contributing to the non-zero acoustic wall admittance and imperfect geometry [29]. The resonance modes of interest for determining speed of sound are the non-degenerate radial modes as they have greater quality factors $Q_{0n} = f_{0n}/(2g)$ than the non-radial modes since they are not influenced by viscous boundary dissipation (motion of the fluid normal to the wall) and are weakly affected by smooth spherical distortion that preserves volume.

The most significant frequency correction at low pressure is due to the thermal boundary layer [5]. The frequency shift $\Delta f_{\mathrm{th}}$ and the contribution to the halfwidth $g_{\mathrm{th}}$ are:

$$\frac{\Delta f_{th}}{f} = -\frac{\gamma - 1}{2a}\delta_{th} + \frac{\gamma - 1}{a} l_{th} + \frac{\gamma - 1}{2a}\delta_{th,w}\frac{\kappa}{\kappa_w} \tag{6}$$

$$\frac{g_{th}}{f} = \frac{\gamma - 1}{2a}\delta_{th} + \frac{\gamma - 1}{2a}\delta_{th,w}\frac{\kappa}{\kappa_w} \tag{7}$$

where the thermal penetration length in the wall $\delta_{\mathrm{th,w}}$ is:

$$\delta_{th,w} = \left[\kappa_w \big/ \left(\pi \rho_w C_{p,w} f\right)\right]^{1/2} \tag{8}$$

with $\rho_{\mathrm{w}}$ the mass density, $\kappa_{\mathrm{w}}$ the thermal conductivity, and $C_{\mathrm{p,w}}$ the isobaric heat capacities of the wall, respectively, and where the thermal accommodation length $l_{\mathrm{th}}$ is:

$$l_{th} = \frac{\kappa}{p}\left(\frac{\pi M T}{2R}\right)^{1/2} \frac{2-h}{h} \frac{1}{C_v M / R + 1/2} \tag{9}$$

with $M$ the molar mass, $R$ the gas constant, $C_{\mathrm{v}}$ the isochoric heat capacity, and $h$ the accommodation coefficient. It is assumed that $h = 1$. This coefficient is dependent on the gas and cavity material and must be determined experimentally, although its value is not significant for speed of sound measurements at the high pressures involved in this work.

The most important frequency correction at high pressures is due to the matching of fluid and resonance cavity velocity in the radial direction [22]. The frequency shift $\Delta f_{\mathrm{sh}}$ is:

$$\Delta f_{sh} = -f\,\frac{\rho w^2}{\rho_w w_w^2}\,q\,\frac{\left(1 + AB - qB^2\right)\tan\left(B - A\right) - \left(B - A\right) - qAB^2}{\left[\left(qA^2 - 1\right)\left(qB^2 - 1\right) + AB\right]\tan\left(B - A\right) - \left(1 + qAB\right)\left(B - A\right)} \tag{10}$$

with:

$$q = \left(1 - \sigma\right) \big/ \left[2\left(1 - 2\sigma\right)\right] \tag{11}$$

$$A = 2\pi fa / w_w \tag{12}$$

$$B = 2\pi fb / w_w \tag{13}$$

where $b$ is the external cavity radius and $\rho_w$ is the density, $w_w$ is the longitudinal speed of sound, and $\sigma$ is the Poisson ratio of the wall material, respectively. This expression is an exact result from elastic theory and is only valid for radial acoustic modes of a spherically perfect cavity. Since the resonance shell is allocated in vacuum there is no contribution to $g$. The elastic properties of steel 321 grade have been approximated to that of steel 304 grade because both stainless steels display similar mechanical behaviour and because more reliable data are found on the latter [30-32]. In principle, acoustic radial modes should only overlap with radial vibrating modes of the cavity, although this is not always true; higher order coupling is possible near the resonance frequencies of the cavity [33]. Acoustic resonance frequencies close to these mechanical resonance frequencies of the cavity are highly perturbed in frequency and halfwidth and appear as outliers relative to the other data. The lower radial symmetric mechanical resonance (breathing frequency) of our shell is estimated to be around $f_{br} \approx 27\cdot10^3$ Hz. Thus, any acoustic resonance frequencies suspected of being close to the mechanical resonance modes of the assembly are discarded.

The inlet gas tube induces a frequency and halfwidth perturbation, $\Delta f_0$ and $g_0$, estimated according to the Kirchhoff-Helmholtz model for closed tubes [21]:

$$\Delta f_0 + i g_0 = \frac{w}{2\pi a}\frac{\Delta S}{4\pi a^2} i y_0 \tag{14}$$

$$y_0 = i \tan\left(k_{KH} L\right) \tag{15}$$

$$k_{KH} = \frac{2\pi f}{w} + \left(1-i\right)\left(\frac{\pi f}{w r_0}\left[\delta_S + \left(\gamma-1\right)\delta_{th}\right]\right) \tag{16}$$

where $\Delta S$ is the section, $r_0$ is the internal radius, and $L$ is the length of the duct, respectively. Two ports are opened in the acoustic cavity: the inlet gas tube of length 80 cm and radius 0.5 mm in the top boss of the cavity, and a blind duct, no longer in use, of length 3.5 cm and radius 0.5 mm in the bottom boss.

Transducers perturbate the resonance by a frequency shift $\Delta f_{tr}$ [5]:

$$\frac{\Delta f_{tr}}{f} = -\frac{\rho w^2 \mathrm{X}_m r_{tr}^2}{2a^3} \tag{17}$$

$$X_m = \left(V_h \kappa_T\right) \Big/ \left(\gamma_{eff} \pi r_{tr}^2\right) \tag{18}$$

where $V_h$ is the volume of the holes drilled in the backplate to increase transducer sensitivity, $r_{tr}$ = 1.5 mm is the transducer radius, $\kappa_T$ is the isothermal compressibility, and $\gamma_{eff}$ = 1 is an effective adiabatic coefficient under the assumption that the gas volume trapped between the dielectric diaphragm and the backplate is small enough to behave as isothermal. The compliance per unit area of our non-commercial transducers is estimated to be $X_m$ = (2.5 to 100) $\cdot 10^{-11}$ m/Pa.

One important effect that produces speed of sound dispersion and absorption at low pressures in some gases, such as methane, is molecular vibrational relaxation. We assume that all the molecules in the mixture relax in unison with a single overall relaxation constant time $\tau_{vib}$, and that excess halfwidth $\Delta g$ is due entirely to the vibrational effect [14], thus:

$$\frac{\Delta g}{f_{0n}} = \frac{\left(g - \left(g_{th} + g_{cl} + g_0\right)\right)}{f_{0n}} = \frac{1}{2}\left(\gamma - 1\right)\Delta 2\pi f \tau_{vib} \tag{19}$$

where $\Delta$ is the vibrational contribution to the isobaric heat capacity of the mixture:

$$\Delta = \sum_k x_K \left(C_{vib,k} / C_p\right) \tag{20}$$

and the molar vibrational heat capacity $C_{vib,k}$ of each pure species $k$ of given composition $x_k$ is estimated from Planck-Einstein functions:

$$C_{vib,k} = R\sum_i \left(z_i^2 e^{z_i}\right) \Big/ \left(e^{z_i} - 1\right)^2 \tag{21}$$

$$z_i = \vartheta_i \big/ T = \frac{h_P \nu_i / k_B}{T} \tag{22}$$

where $h_P$ is Planck's constant and $k_B$ is Boltzmann's constant. For the mixtures in this work, only the molecular vibrational frequencies $\nu_i$ for methane must be considered because helium is monoatomic, and these are taken from spectroscopy data [34]. Frequency correction due to vibrational relaxation $\Delta f_{vib}$ is:

$$\Delta f_{vib} = f\left[\frac{1}{2}(\gamma-1)\Delta(2\pi f\tau_{vib})^2\left(1-\frac{\Delta(1+3\gamma)}{4}\right)\right] \quad (23)$$

Relaxation constant times $\tau_{\text{vib}}$ at $T$ = 273.16 K for (0.95 $CH_4$ + 0.05 He) and (0.90 $CH_4$ + 0.10 He) mixtures are plotted in figure 2. Values increase from (0.05 to 0.5)$\cdot 10^{-6}$ s from the highest to the lowest pressure, with a more pronounced effect for modes (0,5) and (0,6) of higher resonance frequencies, as expected. Similar results are obtained for the other isotherms studied in this work for both mixtures. Note that, the reported experimental values of $\tau_{\text{vib}}$ are just for indication of the magnitudes involved in the vibrational relaxation frequency correction, the goals of this work are beyond the determination of $\tau_{\text{vib}}$ as the function of the inverse of the density for each temperature from the average of the selected acoustic modes. In this way, the incomplete description given by the acoustic model of the real acoustic behaviour of the resonance cavity is not treated as a contribution to the uncertainty of the speed of sound through the addition to $u_r(w_{\text{exp}})$ of the relative excess halfwidths $\Delta g/f$, but it is treated as a perturbation to the resonance frequency under the assumptions described above for the vibrational relaxation phenomena.

The overall frequency corrections, the sum of the effect of the thermal boundary layer, coupling of fluid and shell motion, duct perturbation, transducer perturbation, and vibrational relaxation, take negative values for radial modes (0,2), (0,3), (0,4) and (0,5), which have a resonance frequency below the $f_{\text{br}}$, ranging between (-800 to -100) parts in $10^6$ from the lowest to the highest pressure, and positive values for radial mode (0,6) which has resonance frequencies above $f_{\text{br}}$, ranging from (50 to 500) parts in $10^6$. Thermodynamic and transport property estimations of the working fluid required to calculate the frequency shifts and halfwidths have been taken from open source software CoolProp [35] when the GERG-2008 model was used and RefProp 9.1 software [27] when the AGA8-DC92 model was required. CoolProp implements the GERG mixture model in the same way as RefProp but is open source.

### 3.2. Derived properties.

Experimental speed of sounds obtained from the corrected resonance frequencies are fitted to the acoustic virial equation that can be expressed as a density or pressure series expansion:

$$w^2(p,T) = A_0(T)\left[1+\beta_a(T)\rho+\gamma_a(T)\rho^2+...\right] \quad (24)$$

$$w^2(p,T) = A_0(T)+A_1(T)p+A_2(T)p^2+A_3(T)p^3+A_4(T)p^4+A_5(T)p^5 \quad (25)$$

with:

$$A_0(T) = \frac{RT\gamma^{pg}}{M} = \frac{RT}{M}\frac{C_p^{pg}}{C_p^{pg}-R} \quad (26)$$

$$\beta_a(T) = A_1\frac{M}{\gamma^{pg}} \quad (27)$$

$$\gamma_a - B(T)\beta_a = RTA_2\frac{M}{\gamma^{pg}} \quad (28)$$

where $\beta_a$ and $\gamma_a$ are the second and third acoustic virial coefficients, respectively, and $B(T)$ is the second density virial coefficient. The superscript "*pg*" indicates perfect-gas. The isobaric heat capacity as perfect-gas for a binary mixture of methane and helium $C_{p,mix}^{pg}/R = x_{CH_4}C_{p,CH_4}^{pg}/R + x_{He}C_{p,He}^{pg}/R$ is estimated by the reference AGA8 and GERG-2008 equations of state from the pure helium value $C_{p,He}^{pg}/R = 5/2$ and the expression for the pure methane:

$$\frac{C_{p,CH_4}^{pg}}{R} = B + C\left[\frac{D/T}{\sinh(D/T)}\right]^2 + E\left[\frac{F/T}{\cosh(F/T)}\right]^2 + G\left[\frac{H/T}{\sinh(H/T)}\right]^2 + I\left[\frac{J/T}{\cosh(J/T)}\right]^2 \quad (29)$$

where the regression constants *A* to *J* are obtained by fitting the spectroscopy data of McDowell and Kruse [36] and comparing the speed of sound data of Lemming [13] and Goodwin [37], with good agreement being obtained.

## 4. Results and discussion.

### 4.1. Calibration of the internal radius of the resonance cavity.

Speed of sound measurement requires determining the internal radius of the resonance cavity as a function of temperature and pressure. This task has been accomplished in a previous work [16] by acoustic determination in a fluid of well-known equation of state, such as argon. Data from this previous calibration have been re-analysed to include the same acoustic model for the resonance frequency corrections used in this work. In addition, a new calibration has been performed in argon of purity 99.9999 mol % at $T$ = 300.00 K to check the mechanical stability of the cavity prior to the measurements carried out in this research. The root mean square ($\Delta_{RMS}$) of the relative differences between the old and the new radius calibration is 14 parts in $10^6$ at $T$ = 300 K, which is within the newly determined radius standard relative uncertainty of 97 parts in $10^6$. We thus conclude that, after all the temperature and pressure cycles to which the cavity has been subjected, it is stable enough not to require full recalibration. The internal radius has been estimated from equation (2) applying corrections for the thermal boundary layer, coupling of fluid and shell motion, viscothermal dissipation in the bulk of the fluid, ducts and transducer corrections, and computed from the speed of sound in argon by the reference EoS [39]. Newly fitted coefficients to a polynomial function of pressure for each temperature are given in table 4 together with the radius uncertainty, where the main contribution is due to the expanded uncertainty in speed of sound of 0.02 % of the argon EoS. The truncation order of the polynomial has been chosen in accordance with two criteria: the residuals of the fitting are within the experimental uncertainty and the significance of the parameters obtained from the $p$-value test of statistical significance, which indicates that the uncertainty of the polynomial coefficients does not exceed the value of the coefficient. The deviation of the internal radius from linearity is mainly due to the cavity is not a perfectly isotropic thin-walled spherical shell, instead it is clamped from the north pole boss, the sphere is made by the junction of two hemispheres through a equatorial welded joint, it presents geometrical imperfections caused by the drills for the inlet/outlet gas ducts and transducer plugs, and it is not pressure compensated since it is surrounded by vacuum. Thus, if a linear fit of the determined radius is performed, the residuals overcome the 100 parts in $10^6$ of the argon EoS standard uncertainty, reaching values greater than 110 parts in $10^6$ at the highest pressures and up to 70 parts in $10^6$ at the lowest pressures.

### 4.2. Speed of sound measurements.

Speed of sound data are determined from the average of the (0,2), (0,3) and (0,4) radial acoustic modes, neglecting the (0,5) and (0,6) modes for all the isotherms. As can be seen in figure 3 for $T$ = 273.16 K, the excess halfwidths of the acoustic modes (0,5) and (0,6) are clearly greater than the others over the whole pressure range, indicating that the acoustic model used to analyse the data does not fully describe the acoustic process of the resonance for these modes, as a result of which they are discarded. Relative excess halfwidths for modes (0,2), (0,3), and (0,4) are always well below 50 parts in $10^6$, except for the lowest pressures where the vibrational relaxation effect becomes significant and the frequency correction described in section 3.1 must be applied. Furthermore, relative excess halfwidths from modes (0,5) and (0,6) are above 200 parts in $10^6$ for any pressure at all isotherms. The overall vibrational relaxation times of the mixture $\tau$ have been obtained from the optimisation of the experimental excess halfwidths $\Delta g$ assuming that the vibrational de-excitation of methane molecules in the mixtures obey that $\tau^{-1} = x_{CH4}\cdot\tau_{11}^{-1} + x_{He}\cdot\tau_{12}^{-1}$, where $\tau_{11}$ stands for the vibrational time of pure methane and it is taken from the work of Trusler and Zarari [14], and $\tau_{12}$ stands for the vibrational times associated with unlike collisions, and that the products $\tau_{11}\rho_n$ and $\tau_{12}\rho_n$ are constant along an isotherm, where $\rho_n$ stands for the amount-of-substance density. The values of $\tau_{12}$ derived from these mixtures of ($CH_4$ + He) at 1 kg·m$^{-3}$ are $\tau_{12}$ = (0.535 ± 0.030) μs at $T$ = 273.16 K, $\tau_{12}$ = (0.495 ± 0.030) μs at $T$ = 300 K, $\tau_{12}$ = (0.450 ± 0.027) μs at $T$ = 325 K, $\tau_{12}$ = (0.552 ± 0.037) μs at $T$ = 350 K, and $\tau_{12}$ = (0.498 ± 0.075) μs at $T$ = 375 K. Comparing the experimental relative $\Delta g/f$ for the mixture of higher composition of methane (the relaxing gas) at the highest isotherm, where the greatest values of $\Delta g/f$ are measured, with the $\Delta g/f$ after allowance for vibrational relaxation, it is obtained that the maximum excess halfwidth is reduced from (610 to lower than 140) parts in $10^6$, where the largest correction due to the vibrational relaxation effect is just of 15 parts in $10^6$ for the (0,4) mode at $T$ = 375 K and $p$ = 0.5 MPa. This remaining relative excess halfwidth could be considered into the experimental uncertainty of the speed of sound, which would increase from (230 to 270) parts in $10^6$ at most.

Experimental speed of sound for the two binary ($CH_4$ + He) mixtures studied in this work is shown in tables 5 and 6, together with speed of sound estimated from AGA8-DC92 EoS and GERG-2008 EoS and the relative deviations of the experimental data of this work from the reference equations of state. The results comprise speed of sound data at temperatures $T$ = (273.16, 300, 325, 350, and 375) K and pressures $p$ from (0.5 up to 20) MPa. AGA8-DC92 values have been computed using NIST RefProp 9.1 software [27] and GERG-2008 values using CoolProp software [35]. The expanded relative uncertainty in speed of sound is described in table 7 and draws on contributions from temperature, pressure, gas composition, radius calibration, frequency fitting error to equation (1) and mode dispersion. The overall expanded relative uncertainty ($k$ = 2) in speed of sound is 230 parts in $10^6$. The biggest contribution to this term is due to the uncertainty of the radius calibration from acoustic measurements in argon, which comes from the 200 parts in $10^6$ (0.02 %) expanded uncertainty of argon EoS. Expressions for the uncertainty calculus are detailed in [17].

Square speed of sound data were fitted to the standard virial expansion in pressure given by equation (25). The truncation order is increased until the $\Delta_{RMS}$ of the residuals falls within the average experimental relative uncertainty of the speed of sound. Values of the regression parameters, together with their uncertainties estimated by the Monte Carlo method [38], are shown in table 8. The depth of the concave curve of the speed of sound as a function of pressure for each isotherm decreases as temperature increases. For this reason, a lower polynomial order is required to fit the data within the uncertainty for higher isotherms. It is concluded that a fifth order virial equation for $T$ = 273.16 K, a third order for $T$ = 375 K and a fourth order for the rest of the isotherms are necessary. For example, changing the polynomial regression from fourth to fifth order at the lowest isotherm, $T$ = 273.16 K, means that the $\Delta_{RMS}$ of the residuals decrease from (860 and 240) parts in $10^6$ to (150 and 70) parts in $10^6$ for the (5 and 10) mol-% of hydrogen content mixture, respectively; which are below the experimental uncertainty for the latter case. Figure 4 shows the residuals at each point with no systematic trends and average $\Delta_{RMS}$ of the residuals of (52 and 49) parts in $10^6$ for the (5 and 10) mol-% of He mixtures, respectively, which are five times lower than $U_r(w)$ = 230 parts in $10^6$. This fitting has been done following the same procedure as described in detail in a previous paper [18].

From equations (26) to (28), the adiabatic coefficient $\gamma^{pg}$ and the molar isobaric heat capacity $C_p^{pg}$ are directly derived from speed of sound data in the limit of zero pressure, together with the acoustic second virial coefficient $\beta_a$, and acoustic third virial coefficient $\gamma_a$, for both ($CH_4$ + He) mixtures. Values are reported in table 9, with their corresponding expanded uncertainties ($k$ = 2) and are compared to AGA8-DC92 and GERG-2008 EoS. The expanded experimental relative uncertainty ($k$ = 2) of the derived properties is always better than: 0.02 % for $\gamma^{pg}$, 0.1 % for $C_p^{pg}$, between (0.6 - 8) % for $\beta_a$, and between (1.5 to 6) % for $\gamma_a$. Derived properties as perfect-gas phase are obtained from an extrapolation to zero pressure of the fit performed to the speed of sound data measured in a pressure range from 0.5 to 20 MPa. Although carrying out measurements at lower pressures would improve the extrapolation results, as the pressure is reduced the halfwidth of the resonance lines increases, mainly because of vibrational relaxation phenomena, resulting in a worse fit to equation (1) in the sense of greater resonance frequency uncertainty. For this reason, a limit of 0.5 MPa has been chosen as a compromise between a low enough pressure and good quality acoustic signals.

Figures 5 and 6 show the relative deviations of the measures $w(p,T)$ with respect to the computed values from AGA8-DC92 and GERG-2008 EoS for mixtures (0.95 $CH_4$ + 0.05 He) and (0.90 $CH_4$ + 0.10 He), respectively. Almost all the deviations are outside the expanded experimental uncertainty of 230 parts in $10^6$ (0.023 %), although most of the differences are within the model uncertainty of 2000 parts in $10^6$ (0.2 %) when compared to AGA8 EoS, and roughly half of the results agree with the model uncertainty of 5000 parts in $10^6$ (0.5 %) compared to GERG EoS. In any case, this analysis is highly dependent on composition and temperature. With regard to the models, AGA EoS represents the binary mixtures studied in this work better; the highest relative deviations are between three times lower for mixture (0.95 $CH_4$ + 0.05 He) and four times lower for mixture (0.90 $CH_4$ + 0.10 He) than the differences compared to GERG EoS. In addition, more points are explained within the model uncertainty for AGA EoS than GERG EoS for both compositions, with an absolute average relative deviation ($\Delta_{AAD}$) ranging from (0.25 to 0.45) % for GERG EoS and from (0.09 to 0.14) % for AGA EoS in (0.95 $CH_4$ + 0.05 He) mixture and from (0.48 to 0.83) % for GERG EoS, and from (0.12 to 0.22) % for AGA EoS in (0.90 $CH_4$ + 0.10 He) mixture. As regards the temperature and pressure effect, relative deviations between data and estimated values from models show a nearly linear trend, and relative differences increase with temperature at the lowest pressures and decrease with temperature at the highest, except for the 273.16 K isotherm, which presents a maximum in deviations at an intermediate pressure range and reduced disagreement compared to the two models at both low and high pressures. Experimental data deviate between (-0.15 up to -0.02) % for mixture (0.95 $CH_4$ + 0.05 He) and between (-0.4 up to -0.01) % for mixture (0.90 $CH_4$ + 0.10 He) at low pressures, and relative deviations range from (0.03 up to 0.65) % for mixture (0.95 $CH_4$ + 0.05 He) and from (0.15 up to 1.1) % for mixture (0.90 $CH_4$ + 0.10 He) at high pressures, when compared to AGA EoS. Maximum differences from GERG-2008 EoS are recorded at the lowest isotherm $T$ = 273.16 K and $p$ = (12 to 15) MPa with values exceeding 0.7 % for the 5 % helium mixture and that are as high as 1.3 % for the 10 % helium mixture. The higher the helium content in the gas mixture, the worse the speed of sound estimates are calculated by the models. While for the mixture with a nominal amount of helium of 5 %, nearly all the data are within the model uncertainty, with few discrepancies at pressures between 10 and 15 MPa for the isotherms below 325 K, for the mixture with a nominal molar content of helium of 10 % disagreement is from 7 MPa towards higher pressures at all isotherms, most notably when data are compared to GERG EoS. In general, both models overestimate the value of the speed of sound in these mixtures for pressure below 5 MPa and underestimate it above this point. These findings are summed up as the absolute average relative deviation ($\Delta_{AAD}$), average relative deviation ($\Delta_{Bias}$), root mean square relative deviation ($\Delta_{RMS}$), and maximum relative deviation ($\Delta_{MaxD}$) in table 10.

No data on speed of sound were measured for methane + helium mixtures when this work was carried out. However, the same binary gas samples used in this research were employed in the recent works of Hernández-Gómez et al. [28], [40], where accurate density data measurements were performed at temperatures between (250 and 400) K and pressures up to 20 MPa with a single-sinker densimeter. The results in density obtained by Hernández-Gómez et al. show some similarities with our results in speed of sound when compared to AGA8-DC92 and GERG-2008 EoS. They obtained relative deviations of similar magnitude to that determined by us: their relative differences from AGA8 EoS can also be higher than 0.1 % for mixture (0.95 $CH_4$ + 0.05 He) and exceed the 0.2 % limit of model uncertainty for the mixture (0.90 $CH_4$ + 0.10 He); their relative discrepancies from GERG EoS are even greater in density than in speed of sound, with differences of up to 3 %. As a result, they also conclude that the AGA8 model performs better than the GERG model and that relative deviations increase with the molar content of helium. By contrast, relative deviations in density tend to converge to zero when the pressure is reduced for all the isotherms, while the relative deviations in speed of sound clearly increase with temperature when extrapolating to low pressure for both mixtures. In any case, the differences near zero pressure remain within the AGA and GERG model uncertainty for all the isotherms of both mixtures, apart from the relative deviation at $T$ = 375 K for mixture (0.90 $CH_4$ + 0.10 He).

For the second acoustic virial coefficient $\beta_a$, relative deviations range from (-4 to -31) % for the (0.95 $CH_4$ + 0.05 He) mixture and from (-9 to -140) % for the (0.90 $CH_4$ + 0.10 He) mixture. The values estimated by both models are always greater than the measured data, and discrepancies are similar for both EoS and increase with temperature. For the third acoustic virial coefficient $\gamma_a$, there is no clear trend in the deviations. Except for the case at $T$ = 273.16 K, disagreements are well outside experimental uncertainty. However, the relatively high discrepancies of coefficients $\beta_a$ and $\gamma_a$ when compared to calculations from the AGA and GERG models are to be expected. Both the AGA8-DC92 and GERG-2008 EoS are designed to estimate the thermodynamic properties of pipeline quality natural gas and their application is limited to a range of mole fraction for helium below 0.005, which is far from the molar contents of interest in the helium industry and the nominal concentrations of 5 and 10 % studied in this work. In addition, GERG-2008 EoS only considers vapour liquid equilibrium (VLE) data to fit the binary interactions between methane and helium given that no density, speed of sound, isobaric heat capacity or other caloric or volumetric data of sufficient accuracy were available when this EoS was developed. Thus, binary interactions of methane with helium are only described by adjusted reducing functions of temperature and density. No binary specific departure function or generalized departure function exists for GERG-2008 EoS. For these reasons, the coefficients describing the behaviour of speed of sound with pressure are expected to be correctly obtained by the two models. However, AGA8-DC92 EoS performs better at predicting speed of sound for the mixtures in this work than GERG-2008 EoS. It seems that the formulation of the AGA8-DC92 model which introduces the binary interactions through the second volumetric virial coefficient and the mixture parameter in the reduced density is more suited to predicting thermodynamic properties for mixtures with a molar content outside the model's validity range.

### 4.3. Assessment of mixture stability.

In order to discard the possibility that the disagreements featured in this work are related to a change in the molar mass of the gas filling the resonator during the measurement procedure, the stability of the gas sample was verified with the findings shown in figure 7. Although speed of sound is an intensive thermodynamic property, greater adsorption in the shell wall of one of the mixture components compared to the other might change the molar mass of the mixture and cause systematic deviation in our results. For this reason, continuous measurements of the acoustic (0,3) mode at the pressure of the sample gas bottle under the toughest conditions of the lowest isotherm ($T$ = 273.16 K) and greatest helium content ((0.90 $CH_4$ + 0.10 He) mixture) were recorded for one week, the time required to fully determine each isotherm. The maximum difference in the resonance frequency after six days was 1.7 Hz, which corresponds to a change of 126 parts in $10^6$ and is nearly two times the expanded ($k$ = 2) relative uncertainty contribution of the gas composition to speed of sound uncertainty (table 7). Assuming that helium has been absorbed in a greater proportion than methane in the shell wall and adding this effect to the 230 parts in $10^6$ of speed of sound uncertainty, yields an overall expanded ($k$ = 2) uncertainty of 280 parts in $10^6$, which does not imply any change when discussing the results. A minor effect of the adsorption phenomena is expected at higher temperatures or lower helium content.

## 5. Conclusions.

New speed of sound data for two binary mixtures of (0.95 $CH_4$ + 0.05 He) and (0.90 $CH_4$ + 0.10 He) are reported in the pressure range from 0.5 to 20 MPa at temperatures (273.16, 300, 325, 350, and 375) K with an overall relative expanded ($k$ = 2) uncertainty of 230 parts in $10^6$. These experimental data were fitted to the acoustic virial equation and adiabatic coefficient $\gamma^{pg}$ ($U_r(\gamma^{pg})$ = 0.02 %), and the isobaric heat capacity $C_p^{pg}$ ($U_r(C_p^{pg})$ = 0.1 %), second acoustic virial coefficient $\beta_a$ ($U_r(\beta_a)$ = (0.6 to 8) %) and third acoustic virial coefficient $\gamma_a$ ($U_r(\gamma_a)$ = (1.5 to 6) %) were obtained from the regression parameters.

Speed of sound results were compared to reference models for natural gas-like mixtures: AGA8-DC 92 and GERG-2008 EoS. Relative deviations from model to experimental data are outside experimental uncertainty in most conditions but agree well with model uncertainty: $U_r$(AGA EoS) = 0.2 % and $U_r$(GERG EoS) = 0.5 %. AGA8-DC92 EoS performs better than GERG-2008 EoS when estimating speed of sound according to the data for the ($CH_4$+ He) mixtures in this work. Absolute average deviations are better than 0.45 % for GERG EoS and 0.14 % for AGA EoS in (0.95 $CH_4$ + 0.05 He) mixture, and lower than 0.83 % for GERG EoS and 0.22 % for AGA EoS in (0.90 $CH_4$ + 0.10 He) mixture. Thus, as the molar content of helium increases, poorer agreement is found in the models. The isobaric perfect-gas heat capacities $C_p^{pg}$ derived from speed of sound data present relative deviations greater than 0.2 %, outside the 0.1 % uncertainty of the AGA and GERG models and the experimental one for both ($CH_4$ + He) mixtures, apart from the results at $T$ = 273.16 K, which do agree with the uncertainties. The second acoustic virial coefficient $\beta_a$ differs from models by over 4 % for (0.95 $CH_4$ + 0.05 He) mixture and by over 9 % for (0.90 $CH_4$ + 0.10 He) mixture, with lower values than the EoS estimations in all the isotherms. The third acoustic virial coefficient $\gamma_a$ does not seem to have any specific pattern regarding the relative deviations, with disagreements between (-30 up to +20) %.

This work aims to assess the performance of AGA8-DC92 and GERG-2008 models and to provide new accurate thermodynamic data in the speed of sound domain that can be used to obtain a better correlation of binary ($CH_4$ + He) mixtures, since it has been argued that the two equations of state fail to estimate thermodynamic properties with the accuracy required by industry when the helium content of the mixture is greater than the very low helium amount of substance presented in natural gas-like mixtures.

## Acknowledgements

The authors want to thank for the support to Ministerio de Economía, Industria y Competitividad project ENE2017-88474-R and Junta de Castilla y León project VA280P18.

**Tables and Figures:**

**Table 1.** Mole fraction $x_i$ and expanded ($k$ = 2) uncertainty $U(x_i)$ of the binary methane + helium mixtures studied in this work.

| Composition | ($0.95\ CH_4 + 0.05$ He) | | ($0.90\ CH_4 + 0.10$ He) | |
|---|---|---|---|---|
| | $x_i$ | $U(x_i)$ | $x_i$ | $U(x_i)$ |
| Methane | 0.950015 | 0.000092 | 0.899933 | 0.000083 |
| Helium-4 | 0.049985 | 0.000014 | 0.100067 | 0.000017 |

**Table 2.** Purity, supplier, and critical temperature $T_c$ and pressure $p_c$ of the pure components used for the realization of the binary ($CH_4$ + He) mixtures at BAM. $x$ stands for the mole fraction purity of the pure components.

| Components | Supplier | $x$ | $T_c$ / K(*) | $p_c$ / MPa(*) |
|---|---|---|---|---|
| Methane | Linde AG | ≥ 0.999995 | 190.564 | 4.599 |
| Helium-4 | Linde AG | ≥ 0.999999 | 5.195 | 0.228 |

(*) The critical parameters are computed from RefProp [27].

**Table 3.** Mole fraction $x_i$ and expanded ($k$ = 2) uncertainty $U(x_i)$ from the gas chromatography (GC) analysis, relative deviations of the gravimetric realization from the GC check and gravimetric composition of the validation mixture for the binary ($CH_4$ + He) mixtures. The relative deviations of the gravimetric composition given in Table 1 from the determined composition by the GC analysis are within the stated expanded ($k$ = 2) uncertainty of the composition.

| | ($0.95\ CH_4 + 0.05$ He)<br>BAM nº: 8036-150126 | | | ($0.90\ CH_4 + 0.10$ He)<br>BAM nº: 8069-150127 | | |
|---|---|---|---|---|---|---|
| Components | Composition(*) | | Relative deviation from GC | Composition(*) | | Relative deviation from GC |
| | $x_i \cdot 10^2$ | $U(x_i) \cdot 10^2$ | % | $x_i \cdot 10^2$ | $U(x_i) \cdot 10^2$ | % |
| Methane | 94.796 | 0.031 | -0.22 | 90.019 | 0.040 | 0.03 |
| Helium-4 | 4.9742 | 0.0085 | -0.49 | 10.020 | 0.015 | 0.13 |
| Validation mixture BAM nº: 7065-100105 | | | | | | |
| Components | $x_i \cdot 10^2$ / mol/mol | $U(x_i) \cdot 10^2$ / mol/mol | | | | |
| Methane | 90.4388 | 0.0092 | | | | |
| Helium-4 | 9.5599 | 0.0060 | | | | |
| Carbon Monoxide | 0.0002158 | 0.0000002 | | | | |
| Carbon Dioxide | 0.0002164 | 0.0000002 | | | | |
| Oxygen | 0.0002139 | 0.0000002 | | | | |
| Argon | 0.0002169 | 0.0000002 | | | | |
| Hydrogen | 0.0002220 | 0.0000003 | | | | |
| Nitrogen | 0.0002166 | 0.0000002 | | | | |

(*) The mole fractions specified in this table are not normalized, thus their sum is not equal to 1.

**Table 4.** Regression coefficients of the internal resonance cavity radius $a$ and expanded ($k$ = 2) relative uncertainties $U_r(a)$ to the pressure function: $a = a_0 + a_1 \cdot p + a_2 \cdot p^2$.

| $T$ / K | $10^2 \cdot a_0$ / m | $10^7 \cdot a_1$ / m·MPa$^{-1}$ | $10^8 \cdot a_2$ / m·MPa$^{-2}$ | $10^6 \cdot U_r(a)$ / m·s$^{-1}$/ m·s$^{-1}$ |
|---|---|---|---|---|
| 273.16 | 4.016094 | 8.327 | 6.458 | 190 |
| 300.00 | 4.017802 | 8.471 | 5.582 | 200 |
| 300.00* | 4.017757 | 9.166 | 5.051 | 170 |
| 325.00 | 4.019559 | 13.145 | 3.163 | 220 |
| 350.00 | 4.020978 | 14.646 | 2.090 | 220 |
| 375.00 | 4.022621 | 15.378 | 1.870 | 240 |

(*) Test measurement.

**Table 5.** Experimental speeds of sound $w_{exp}$ with their corresponding relative expanded ($k$ = 2) uncertainties[(*)] after applying the acoustic model and data reduction, and comparison with EoS GERG-2008 and AGA8-DC92 for (0.95 $CH_4$ + 0.05 He) mixture with the composition specified in Table 1. Also, the average of the ratio between the experimental resonance frequencies $f_{0n}$ directly measured and the eigenvalues $\nu_{0n}$ are reported.

| $p$ / MPa | $\langle f_{0n}/\nu_{0n}\rangle$ (**) / Hz | $w_{exp}$ / m·s$^{-1}$ | $10^2\cdot(w_{exp} - w_{AGA})/w_{AGA}$ | $10^2\cdot(w_{exp} - w_{GERG})/w_{GERG}$ | $p$ / MPa | $\langle f_{0n}/\nu_{0n}\rangle$ (**) / Hz | $w_{exp}$ / m·s$^{-1}$ | $10^2\cdot(w_{exp} - w_{AGA})/w_{AGA}$ | $10^2\cdot(w_{exp} - w_{GERG})/w_{GERG}$ |
|---|---|---|---|---|---|---|---|---|---|
| | | $T$ = 273.16 K | | | | | $T$ = 300.00 K | | |
| 0.49533 | 1739.127 | 438.872 | -0.031 | -0.023 | 0.48364 | 1817.060 | 458.745 | -0.069 | -0.062 |
| 1.01024 | 1732.128 | 437.114 | -0.004 | 0.012 | 0.98569 | 1812.872 | 457.694 | -0.052 | -0.035 |
| 2.01057 | 1719.311 | 433.899 | 0.021 | 0.061 | 1.99598 | 1805.321 | 455.802 | -0.030 | 0.007 |
| 2.98884 | 1708.593 | 431.215 | 0.045 | 0.116 | 3.21273 | 1798.487 | 454.098 | -0.005 | 0.063 |
| 4.42328 | 1696.837 | 428.279 | 0.079 | 0.211 | 4.47517 | 1794.394 | 453.091 | 0.018 | 0.130 |
| 5.98410 | 1690.792 | 426.789 | 0.117 | 0.332 | 5.51326 | 1793.763 | 452.955 | 0.045 | 0.198 |
| 6.47585 | 1690.604 | 426.753 | 0.127 | 0.371 | 6.46078 | 1795.348 | 453.379 | 0.064 | 0.256 |
| 6.97239 | 1691.367 | 426.958 | 0.140 | 0.412 | 7.01185 | 1797.351 | 453.899 | 0.075 | 0.291 |
| 7.52433 | 1693.359 | 427.475 | 0.153 | 0.456 | 7.54296 | 1800.052 | 454.596 | 0.085 | 0.323 |
| 8.58946 | 1701.048 | 429.444 | 0.183 | 0.541 | 8.43077 | 1806.420 | 456.229 | 0.105 | 0.380 |
| 9.52381 | 1712.101 | 432.263 | 0.202 | 0.604 | 9.34419 | 1815.340 | 458.509 | 0.122 | 0.431 |
| 10.52348 | 1728.759 | 436.501 | 0.218 | 0.661 | 10.53354 | 1831.862 | 462.721 | 0.195 | 0.547 |
| 11.50605 | 1750.170 | 441.941 | 0.227 | 0.701 | 11.49983 | 1847.753 | 466.769 | 0.206 | 0.589 |
| 12.53237 | 1778.002 | 449.010 | 0.234 | 0.726 | 12.50669 | 1867.470 | 471.789 | 0.212 | 0.624 |
| 13.31689 | 1802.841 | 455.316 | 0.228 | 0.723 | 13.30529 | 1885.404 | 476.352 | 0.215 | 0.646 |
| 14.04039 | 1828.424 | 461.811 | 0.219 | 0.705 | 13.99831 | 1902.557 | 480.716 | 0.212 | 0.659 |
| 14.72231 | 1854.704 | 468.484 | 0.206 | 0.673 | 14.69891 | 1921.389 | 485.507 | 0.210 | 0.668 |
| 15.29918 | 1878.519 | 474.531 | 0.194 | 0.639 | 15.30186 | 1938.723 | 489.916 | 0.202 | 0.670 |
| 15.85321 | 1902.602 | 480.648 | 0.181 | 0.599 | 15.90160 | 1956.971 | 494.558 | 0.194 | 0.668 |
| 16.51590 | 1932.887 | 488.342 | 0.162 | 0.548 | 16.49104 | 1975.844 | 499.360 | 0.185 | 0.662 |
| 17.03223 | 1957.426 | 494.578 | 0.147 | 0.504 | 17.00462 | 1993.059 | 503.740 | 0.179 | 0.656 |
| 17.53263 | 1982.025 | 500.832 | 0.134 | 0.464 | 17.52499 | 2011.417 | 508.329 | 0.167 | 0.644 |
| 18.00923 | 2005.955 | 506.918 | 0.118 | 0.421 | 18.01263 | 2028.909 | 512.777 | 0.158 | 0.631 |
| 18.55186 | 2034.352 | 514.011 | 0.098 | 0.372 | 18.51484 | 2047.441 | 517.493 | 0.148 | 0.615 |
| 19.08831 | 2062.626 | 521.156 | 0.074 | 0.322 | 19.01097 | 2066.285 | 522.290 | 0.140 | 0.600 |
| 19.87071 | 2104.748 | 531.762 | 0.032 | 0.243 | 19.44713 | 2083.212 | 526.603 | 0.132 | 0.585 |
| | | *T* = 325.00 K | | | | | *T* = 350.00 K | | |
| 0.47957 | 1884.782 | 476.043 | -0.060 | -0.053 | 0.47677 | 1946.841 | 491.904 | -0.117 | -0.112 |
| 0.97932 | 1882.400 | 475.453 | -0.044 | -0.031 | 0.98581 | 1945.915 | 491.679 | -0.100 | -0.089 |
| 1.67736 | 1879.596 | 474.762 | -0.025 | -0.001 | 1.98848 | 1944.771 | 491.412 | -0.075 | -0.052 |
| 2.45546 | 1877.248 | 474.189 | -0.007 | 0.033 | 3.55771 | 1945.839 | 491.722 | -0.026 | 0.026 |
| 3.48142 | 1875.723 | 473.831 | 0.023 | 0.089 | 5.00545 | 1949.467 | 492.679 | 0.000 | 0.087 |
| 4.48607 | 1875.709 | 473.854 | 0.045 | 0.138 | 6.52087 | 1956.582 | 494.522 | 0.027 | 0.155 |
| 5.51015 | 1877.536 | 474.344 | 0.066 | 0.191 | 8.53732 | 1971.525 | 498.362 | 0.057 | 0.241 |
| 6.49807 | 1881.157 | 475.287 | 0.086 | 0.244 | 10.53433 | 1992.879 | 503.830 | 0.085 | 0.323 |
| 7.00128 | 1883.758 | 475.958 | 0.096 | 0.271 | 12.03182 | 2013.191 | 509.021 | 0.100 | 0.374 |
| 7.49852 | 1886.818 | 476.746 | 0.106 | 0.298 | 13.41811 | 2036.102 | 514.869 | 0.152 | 0.454 |
| 8.51246 | 1894.633 | 478.751 | 0.126 | 0.351 | 14.51942 | 2055.933 | 519.931 | 0.158 | 0.484 |
| 9.50390 | 1904.389 | 481.247 | 0.141 | 0.398 | 15.59913 | 2077.205 | 525.360 | 0.160 | 0.509 |
| 10.50173 | 1916.395 | 484.313 | 0.156 | 0.443 | 16.62592 | 2099.045 | 530.933 | 0.159 | 0.527 |
| 11.46917 | 1930.127 | 487.817 | 0.168 | 0.481 | 17.51164 | 2119.089 | 536.049 | 0.155 | 0.539 |
| 12.51699 | 1947.513 | 492.249 | 0.183 | 0.525 | 18.43731 | 2141.261 | 541.708 | 0.151 | 0.552 |
| 13.29503 | 1961.870 | 495.908 | 0.187 | 0.547 | 19.23698 | 2161.235 | 546.808 | 0.142 | 0.557 |

| | | | | |
|---|---|---|---|---|
| 14.01107 | 1976.478 | 499.577 | 0.190 | 0.566 |
| 14.71621 | 1991.742 | 503.462 | 0.189 | 0.582 |
| 15.30860 | 2005.392 | 506.936 | 0.189 | 0.594 |
| 16.00260 | 2022.254 | 511.232 | 0.187 | 0.605 |
| 16.55219 | 2036.316 | 514.814 | 0.186 | 0.613 |
| 17.00200 | 2048.236 | 517.854 | 0.183 | 0.619 |
| 17.51314 | 2062.248 | 521.429 | 0.180 | 0.624 |
| 17.96077 | 2074.665 | 524.657 | 0.179 | 0.628 |
| 18.52073 | 2091.313 | 528.846 | 0.179 | 0.636 |
| 19.02448 | 2106.392 | 532.708 | 0.176 | 0.638 |
| 19.53617 | 2122.175 | 536.749 | 0.175 | 0.641 |
| | | *T* = 375.00 K | | |
| 0.50808 | 2005.340 | 506.893 | -0.163 | -0.160 |
| 0.97030 | 2005.510 | 506.945 | -0.147 | -0.141 |
| 1.99494 | 2006.522 | 507.226 | -0.119 | -0.103 |
| 3.48611 | 2009.864 | 508.112 | -0.090 | -0.053 |
| 4.99434 | 2015.884 | 509.679 | -0.063 | 0.004 |
| 6.50740 | 2024.642 | 511.941 | -0.037 | 0.062 |
| 8.53451 | 2041.381 | 516.241 | 0.018 | 0.165 |
| 10.52641 | 2062.461 | 521.645 | 0.045 | 0.237 |
| 12.04093 | 2081.959 | 526.636 | 0.064 | 0.289 |
| 13.43239 | 2102.405 | 531.867 | 0.075 | 0.329 |
| 14.50969 | 2119.943 | 536.352 | 0.084 | 0.359 |
| 15.60939 | 2139.303 | 541.303 | 0.091 | 0.387 |
| 16.57330 | 2157.411 | 545.933 | 0.094 | 0.407 |
| 20.02005 | 2181.545 | 551.995 | 0.133 | 0.559 |
| 20.96589 | 2207.811 | 558.461 | 0.111 | 0.551 |
| 17.51813 | 2176.110 | 550.715 | 0.092 | 0.421 |
| 18.32191 | 2193.569 | 555.023 | 0.096 | 0.440 |
| 19.19812 | 2212.675 | 559.901 | 0.097 | 0.456 |

(*) Expanded uncertainties ($k$ = 2): $U(p) = 7.5 \cdot 10^{-5}$ ($p$/Pa) + 200 Pa; $U(T)$ = 4 mK; $U_r(w) = 2.3 \cdot 10^{-4}$ m·s$^{-1}$/ m·s$^{-1}$.

(**) Average of the experimental measured frequencies before applying the acoustic model.

**Table 6.** Experimental speeds of sound $w_{exp}$ with their corresponding relative expanded ($k$ = 2) uncertainties(*) after applying the acoustic model and data reduction, and comparison with EoS GERG-2008 and AGA8-DC92 for (0.90 $CH_4$ + 0.10 He) mixture with the composition specified in Table 1. Also, the average of the ratio between the experimental resonance frequencies $f_{0n}$ directly measured and the eigenvalues $\nu_{0n}$ are reported.

| $p$ / MPa | $\langle f_{0n}/\nu_{0n}\rangle$ (**) / Hz | $w_{exp}$ / m·s$^{-1}$ | $10^2\cdot(w_{exp} - w_{AGA})/w_{AGA}$ | $10^2\cdot(w_{exp} - w_{GERG})/w_{GERG}$ | $p$ / MPa | $\langle f_{0n}/\nu_{0n}\rangle$ (**) / Hz | $w_{exp}$ / m·s$^{-1}$ | $10^2\cdot(w_{exp} - w_{AGA})/w_{AGA}$ | $10^2\cdot(w_{exp} - w_{GERG})/w_{GERG}$ |
|---|---|---|---|---|---|---|---|---|---|
| | | $T$ = 273.16 K | | | | | $T$ = 300.00 K | | |
| 0.49107 | 1782.417 | 449.799 | -0.006 | 0.012 | 0.48233 | 1861.034 | 469.850 | -0.086 | -0.069 |
| 0.99340 | 1777.233 | 448.499 | 0.019 | 0.059 | 1.00098 | 1858.342 | 469.176 | -0.054 | -0.017 |
| 1.70533 | 1770.523 | 446.820 | 0.050 | 0.126 | 1.98554 | 1853.866 | 468.062 | -0.019 | 0.064 |
| 2.48493 | 1764.215 | 445.245 | 0.082 | 0.207 | 2.98462 | 1850.949 | 467.344 | 0.017 | 0.155 |
| 3.49042 | 1757.889 | 443.670 | 0.122 | 0.321 | 4.48726 | 1849.857 | 467.101 | 0.065 | 0.303 |
| 4.49837 | 1753.895 | 442.685 | 0.160 | 0.447 | 5.49005 | 1851.653 | 467.577 | 0.097 | 0.411 |
| 5.50571 | 1752.571 | 442.375 | 0.196 | 0.582 | 6.49483 | 1855.564 | 468.590 | 0.126 | 0.518 |
| 6.49302 | 1754.163 | 442.801 | 0.230 | 0.717 | 7.50143 | 1861.835 | 470.201 | 0.156 | 0.627 |
| 7.50809 | 1759.070 | 444.067 | 0.261 | 0.852 | 9.02642 | 1875.758 | 473.764 | 0.185 | 0.771 |
| 8.48165 | 1767.396 | 446.196 | 0.303 | 0.987 | 10.50357 | 1895.098 | 478.718 | 0.226 | 0.910 |
| 9.52113 | 1779.719 | 449.338 | 0.317 | 1.089 | 11.50501 | 1911.324 | 482.859 | 0.244 | 0.985 |
| 10.52230 | 1795.753 | 453.420 | 0.337 | 1.182 | 12.49205 | 1929.875 | 487.560 | 0.252 | 1.043 |
| 11.53033 | 1815.878 | 458.538 | 0.350 | 1.253 | 13.50538 | 1951.570 | 493.085 | 0.266 | 1.100 |
| 12.63771 | 1842.608 | 465.332 | 0.355 | 1.300 | 14.49513 | 1974.989 | 499.048 | 0.263 | 1.132 |
| 13.63337 | 1870.622 | 472.451 | 0.348 | 1.308 | 15.51452 | 2001.753 | 505.799 | 0.254 | 1.149 |
| 14.53678 | 1899.154 | 479.702 | 0.334 | 1.284 | 16.48981 | 2029.034 | 512.809 | 0.241 | 1.151 |
| 15.35091 | 1926.998 | 486.780 | 0.303 | 1.225 | 16.99359 | 2044.055 | 516.634 | 0.233 | 1.147 |
| 16.04141 | 1952.616 | 493.253 | 0.283 | 1.169 | 17.51091 | 2060.338 | 520.691 | 0.223 | 1.138 |
| 16.53703 | 1971.771 | 498.116 | 0.267 | 1.121 | 18.01162 | 2076.295 | 524.748 | 0.214 | 1.127 |
| 17.04809 | 1992.221 | 503.308 | 0.249 | 1.065 | 18.52305 | 2093.007 | 529.002 | 0.202 | 1.110 |
| 17.54776 | 2012.861 | 508.548 | 0.230 | 1.007 | 19.04285 | 2110.475 | 533.448 | 0.190 | 1.091 |
| 18.05910 | 2034.220 | 514.054 | 0.208 | 0.943 | 19.95515 | 2142.046 | 541.498 | 0.164 | 1.046 |
| 18.53662 | 2054.909 | 519.322 | 0.186 | 0.881 | | | | | |
| 19.11694 | 2080.647 | 525.878 | 0.158 | 0.805 | | | | | |
| | | $T$ = 325.00 K | | | | | $T$ = 350.00 K | | |
| 0.47390 | 1928.142 | 486.997 | -0.179 | -0.164 | 0.48940 | 1990.910 | 503.045 | -0.259 | -0.246 |
| 0.97685 | 1927.456 | 486.836 | -0.133 | -0.101 | 0.99000 | 1991.081 | 503.092 | -0.235 | -0.208 |
| 1.47754 | 1926.719 | 486.663 | -0.107 | -0.056 | 1.98152 | 1992.168 | 503.391 | -0.192 | -0.132 |
| 2.48156 | 1925.935 | 486.492 | -0.073 | 0.021 | 3.47680 | 1995.654 | 504.312 | -0.148 | -0.028 |
| 3.48642 | 1926.667 | 486.705 | -0.037 | 0.108 | 4.99327 | 2002.607 | 506.115 | -0.080 | 0.113 |
| 4.48552 | 1928.829 | 487.280 | -0.006 | 0.194 | 6.48403 | 2011.896 | 508.509 | -0.041 | 0.230 |
| 5.48725 | 1932.684 | 488.283 | 0.027 | 0.289 | 7.48758 | 2020.022 | 510.596 | -0.009 | 0.316 |
| 6.48473 | 1938.091 | 489.677 | 0.054 | 0.379 | 8.49370 | 2029.446 | 513.012 | 0.016 | 0.395 |
| 7.48333 | 1945.327 | 491.535 | 0.083 | 0.473 | 9.50208 | 2040.443 | 515.827 | 0.044 | 0.476 |
| 8.49739 | 1954.419 | 493.863 | 0.108 | 0.561 | 10.50314 | 2052.654 | 518.950 | 0.064 | 0.546 |
| 9.49423 | 1965.277 | 496.639 | 0.134 | 0.648 | 11.99627 | 2073.725 | 524.334 | 0.100 | 0.651 |

| | | | | |
|---|---|---|---|---|
| 10.49266 | 1977.818 | 499.871 | 0.154 | 0.725 |
| 11.47531 | 1992.361 | 503.566 | 0.183 | 0.805 |
| 12.50025 | 2009.223 | 507.854 | 0.198 | 0.870 |
| 13.50338 | 2027.528 | 512.522 | 0.210 | 0.925 |
| 14.49345 | 2047.276 | 517.559 | 0.214 | 0.967 |
| 15.29358 | 2064.882 | 521.983 | 0.222 | 1.003 |
| 15.99451 | 2080.885 | 526.061 | 0.223 | 1.027 |
| 16.48464 | 2092.634 | 529.048 | 0.225 | 1.043 |
| 16.99220 | 2105.127 | 532.230 | 0.223 | 1.056 |
| 17.52021 | 2118.597 | 535.657 | 0.223 | 1.069 |
| 17.99545 | 2130.974 | 538.814 | 0.218 | 1.075 |
| 18.49030 | 2144.523 | 542.215 | 0.218 | 1.085 |
| 18.99341 | 2157.632 | 545.758 | 0.216 | 1.092 |
| 19.50394 | 2172.078 | 549.411 | 0.208 | 1.091 |
| 19.90577 | 2183.714 | 552.384 | 0.208 | 1.097 |
| | | $T$ = 375.00 K | | |
| 0.48089 | 2047.906 | 517.606 | -0.444 | -0.435 |
| 0.97452 | 2049.437 | 518.004 | -0.399 | -0.378 |
| 1.97777 | 2052.166 | 518.720 | -0.363 | -0.315 |
| 2.98245 | 2056.242 | 519.779 | -0.314 | -0.233 |
| 4.47476 | 2063.604 | 521.686 | -0.275 | -0.137 |
| 5.99390 | 2074.533 | 524.499 | -0.191 | 0.012 |
| 7.48290 | 2086.835 | 527.662 | -0.154 | 0.116 |
| 8.48587 | 2097.104 | 530.296 | -0.104 | 0.213 |
| 9.49486 | 2108.291 | 533.164 | -0.068 | 0.295 |
| 13.49805 | 2097.746 | 530.471 | 0.116 | 0.729 |
| 14.49821 | 2115.639 | 535.040 | 0.133 | 0.783 |
| 15.50500 | 2135.197 | 539.958 | 0.142 | 0.829 |
| 16.49997 | 2155.638 | 545.162 | 0.152 | 0.872 |
| 17.49527 | 2177.106 | 550.647 | 0.155 | 0.904 |
| 18.50851 | 2200.325 | 556.568 | 0.160 | 0.939 |
| 19.50930 | 2224.170 | 562.670 | 0.158 | 0.963 |
| 20.35239 | 2245.223 | 568.049 | 0.160 | 0.985 |
| 10.49061 | 2120.219 | 536.220 | -0.045 | 0.362 |
| 11.48169 | 2133.444 | 539.605 | -0.012 | 0.436 |
| 12.49983 | 2148.529 | 543.464 | 0.036 | 0.526 |
| 13.46461 | 2163.865 | 547.302 | 0.065 | 0.592 |
| 14.47474 | 2181.565 | 551.755 | 0.122 | 0.686 |
| 15.49262 | 2199.769 | 556.408 | 0.158 | 0.756 |
| 16.43996 | 2217.278 | 560.926 | 0.179 | 0.808 |
| 17.48972 | 2238.484 | 566.340 | 0.226 | 0.887 |
| 18.17166 | 2252.268 | 569.916 | 0.240 | 0.922 |
| 18.70492 | 2263.639 | 572.844 | 0.261 | 0.958 |

(*) Expanded uncertainties ($k$ = 2): $U(p) = 7.5 \cdot 10^{-5}$ ($p$/Pa) + 200 Pa; $U(T)$ = 4 mK; $U_r(w) = 2.3 \cdot 10^{-4}$ m·s$^{-1}$/ m·s$^{-1}$.

(**) Average of the experimental measured frequencies before applying the acoustic model.

**Table 7.** Uncertainty budget for the speed of sound $w$ measurements. Unless otherwise specified, uncertainty $u$ is indicated with a coverage factor $k$ = 1.

| Source | Magnitude | | Contribution to speed of sound uncertainty, $10^6 \cdot u_r(w)$ / (m·s$^{-1}$)/(m·s$^{-1}$) |
|---|---|---|---|
| *State-point uncertainties* | | | |
| Temperature | Calibration | 0.002 K | |
| | Resolution | $7.2 \cdot 10^{-7}$ K | |
| | Repeatability | $5.6 \cdot 10^{-5}$ K | |
| | Gradient (across hemispheres) | $4.0 \cdot 10^{-3}$ K | |
| | Sum | 0.005 K | 5.0 |
| Pressure | Calibration | $(7.5 \cdot 10^{-5} \cdot p + 2 \cdot 10^{-4})$ MPa | |
| | Resolution | $2.9 \cdot 10^{-5}$ MPa | |
| | Repeatability | $1.22 \cdot 10^{-5}$ MPa | |
| | Sum | $(1.2 \text{ to } 8.4) \cdot 10^{-4}$ MPa | 5.1 |
| Gas composition | Purity | $8.3\cdot10^{-7}$ kg/mol | |
| | Molar mass | $7.0\cdot10^{-7}$ kg/mol | |
| | Sum | $1.1\cdot10^{-6}$ kg/mol | 34.8 |
| *Cavity radius* | | | |
| Radius from speed of sound in Ar | Temperature | $1.5\cdot10^{-9}$ m | |
| | Pressure | $1.6\cdot10^{-10}$ m | |
| | Gas Composition | $4.1\cdot10^{-9}$ m | |
| | Frequency fitting | $4.9\cdot10^{-7}$ m | |
| | Regression | $1.7\cdot10^{-6}$ m | |
| | Equation of State | $2.3\cdot10^{-6}$ m | |
| | Dispersion of modes | $2.9\cdot10^{-6}$ m | |
| | Sum | $4.2\cdot10^{-6}$ m | 105.7 |
| *Fitting of, and corrections to, resonance frequency* | | | |
| Frequency fitting | | 0.0012 Hz | 2.6 |
| Dispersion of modes | | $1.8\cdot10^{-2}$ m·s$^{-1}$ | 46.2 |
| Sum of all contributions to *w* | | | 117.0 |
| $10^6\cdot U_r(w)$ / (m·s$^{-1}$)/(m·s$^{-1}$) (*) | | | 233.9 |

(*) Uncertainty with coverage factor $k = 2$.

**Table 8.** Fitting parameters $A_i(T)$ of the square speed of sound according equation (25) and their corresponding expanded ($k = 2$) uncertainties determined by the MonteCarlo method, including a comparison of the experimental $A_{0,\,exp}$ with the theoretical $A_{0,\,calc}$ calculated from the known properties of the pure components and the gas composition according to equations (26) and (29).

| $T$ / K | $A_0(T)$ / m$^2$·s$^{-2}$ | $A_1(T)$ / m$^2$·s$^{-2}$·Pa$^{-1}$ | $A_2(T)$ / m$^2$·s$^{-2}$·Pa$^{-2}$ | $A_3(T)$ / m$^2$·s$^{-2}$·Pa$^{-3}$ | $A_4(T)$ / m$^2$·s$^{-2}$·Pa$^{-4}$ | $A_5(T)$ / m$^2$·s$^{-2}$·Pa$^{-5}$ | $\Delta_{RMS}$ of the residuals / ppm | ($A_{0,exp}$ - $A_{0,calc}$)/$A_{0,calc}$ / % |
|---|---|---|---|---|---|---|---|---|
| | | | | (0.95 $CH_4$ + 0.05 He) | | | | |
| 273.16 | 194207 ± 23 | (-328.6 ± 3.0)·$10^{-5}$ | (15.3 ± 1.0)·$10^{-11}$ | (5.1 ± 1.5)·$10^{-18}$ | (101.5 ± 8.7)·$10^{-26}$ | (-33.8 ± 1.9)·$10^{-33}$ | 44 | -0.094 |
| 300 | 211401 ± 19 | (-205.7 ± 1.4)·$10^{-5}$ | (94.7 ± 3.0)·$10^{-12}$ | (144.6 ± 2.3)·$10^{-19}$ | (-253.7 ± 5.8)·$10^{-27}$ | - | 71 | -0.20 |
| 325 | 227180 ± 20 | (-127.4.0 ± 1.5)·$10^{-5}$ | (119.7 ± 3.2)·$10^{-12}$ | (78.7 ± 2.5)·$10^{-19}$ | (-127.5 ± 6.2)·$10^{-27}$ | - | 24 | -0.17 |
| 350 | 242197 ± 33 | $(-56.9 \pm 2.6)\cdot10^{-5}$ | $(105.7 \pm 5.6)\cdot10^{-12}$ | $(63.5 \pm 4.2)\cdot10^{-19}$ | $(-12.1 \pm 1.0)\cdot10^{-26}$ | - | 63 | -0.29 |
| 375 | 257061 ± 28 | $(-21.4 \pm 1.5)\cdot10^{-5}$ | $(146.7 \pm 1.8)\cdot10^{-12}$ | $(92.1 \pm 6.4)\cdot10^{-20}$ | - | - | 59 | -0.32 |
| | | | | ($0.90\ CH_4 + 0.10\ He$) | | | | |
| 273.16 | 203527 ± 20 | $(-254.5 \pm 2.3)\cdot10^{-5}$ | $(143.5 \pm 7.4)\cdot10^{-12}$ | $(98.6 \pm 9.9)\cdot10^{-19}$ | $(26.6 \pm 5.7)\cdot10^{-26}$ | $(-13.8 \pm 1.2)\cdot10^{-33}$ | 33 | -0.071 |
| 300 | 221415 ± 21 | $(-142.9 \pm 1.6)\cdot10^{-5}$ | $(112.8 \pm 3.4)\cdot10^{-12}$ | $(110.6 \pm 2.6)\cdot10^{-19}$ | $(-204.6 \pm 6.6)\cdot10^{-27}$ | - | 33 | -0.24 |
| 325 | 237511 ± 20 | $(-67.3 \pm 1.6)\cdot10^{-5}$ | $(119.8 \pm 3.3)\cdot10^{-12}$ | $(67.0 \pm 2.5)\cdot10^{-19}$ | $(-122.8 \pm 6.4)\cdot10^{-27}$ | - | 38 | -0.39 |
| 350 | 253076 ± 32 | $(-10.8 \pm 2.5)\cdot10^{-5}$ | $(127.7 \pm 5.2)\cdot10^{-12}$ | $(34.8 \pm 3.8)\cdot10^{-19}$ | $(-60.7 \pm 9.2)\cdot10^{-27}$ | - | 34 | -0.56 |
| 375 | 267796 ± 30 | $(36.2 \pm 1.6)\cdot10^{-5}$ | $(135.1 \pm 2.1)\cdot10^{-12}$ | $(97.5 \pm 7.4)\cdot10^{-20}$ | - | - | 107 | -0.89 |

**Table 9.** Adiabatic coefficient $\gamma^{pg}$, isobaric heat capacity $C_p^{pg}$, acoustic second virial coefficient $\beta_a$, and acoustic third virial coefficient $\gamma_a$ derived from the speed of sound data with their corresponding relative expanded ($k$ = 2) uncertainty $U_r$ and comparison with AGA8 and GERG-2008 EoS. The superscript *pg* indicates perfect-gas property.

| $T$ / K | $\gamma^{pg}$ | $10^2\cdot U_r(\gamma^{pg})$ | $10^2\cdot(\gamma^{pg}_{exp} - \gamma^{pg}_{GERG})/\gamma^{pg}_{GERG}$ | $10^2\cdot(\gamma^{pg}_{exp} - \gamma^{pg}_{AGA})/\gamma^{pg}_{AGA}$ | $C_p^{pg}$ / $J\cdot mol^{-1}\cdot K^{-1}$ | $10^2\cdot U_r(C_p^{pg})$ | $10^2\cdot(C_p^{pg}{}_{exp} - C_p^{pg}{}_{GERG})/C_p^{pg}{}_{GERG}$ | $10^2\cdot(C_p^{pg}{}_{exp} - C_p^{pg}{}_{AGA})/C_p^{pg}{}_{AGA}$ |
|---|---|---|---|---|---|---|---|---|
| | | | | ($0.95\ CH_4 + 0.05\ He$) | | | | |
| 273.16 | 1.32035 | 0.018 | -0.091 | -0.095 | 34.269 | 0.082 | 0.30 | 0.29 |
| 300.00 | 1.30866 | 0.017 | -0.19 | -0.20 | 35.252 | 0.074 | 0.64 | 0.64 |
| 325.00 | 1.29816 | 0.018 | -0.16 | -0.17 | 36.201 | 0.077 | 0.57 | 0.57 |
| 350.00 | 1.28512 | 0.020 | -0.29 | -0.29 | 37.476 | 0.091 | 1.0 | 1.0 |
| 375.00 | 1.27305 | 0.019 | -0.32 | -0.31 | 38.765 | 0.088 | 1.2 | 1.2 |
| | | | | ($0.90\ CH_4 + 0.10\ He$) | | | | |
| 273.16 | 1.32967 | 0.018 | -0.071 | -0.071 | 33.535 | 0.072 | 0.22 | 0.21 |
| 300.00 | 1.31712 | 0.017 | -0.24 | -0.24 | 34.533 | 0.075 | 0.75 | 0.75 |
| 325.00 | 1.30419 | 0.017 | -0.39 | -0.39 | 35.647 | 0.073 | 1.3 | 1.3 |
| 350.00 | 1.29040 | 0.019 | -0.56 | -0.56 | 36.946 | 0.087 | 1.9 | 1.9 |
| 375.00 | 1.27442 | 0.019 | -0.89 | -0.89 | 38.613 | 0.088 | 3.3 | 3.3 |

| $T$ | $\beta_a$ / $m^3 \cdot mol^{-1}$ | $10^2 \cdot U_r(\beta_a)$ | $10^2 \cdot (\beta_{a,exp} - \beta_{a,GERG})/\beta_{a,GERG}$ | $10^2 \cdot (\beta_{a,exp} - \beta_{a,AGA})/\beta_{a,AGA}$ | $\gamma_a$ / $(m^3 \cdot mol^{-1})^2$ | $10^2 \cdot U_r(\gamma_a)$ | $10^2 \cdot (\gamma_{a,exp} - \gamma_{a,GERG})/\gamma_{a,GERG}$ | $10^2 \cdot (\gamma_{a,exp} - \gamma_{a,AGA})/\gamma_{a,AGA}$ |
|---|---|---|---|---|---|---|---|---|
| | | | | ($0.95\ CH_4 + 0.05\ He$) | | | | |
| 273.16 | $-384.3 \cdot 10^{-7}$ | 0.88 | -6.0 | -4.3 | $58.4 \cdot 10^{-10}$ | 4.8 | -3.0 | -3.4 |
| 300.00 | $-242.7 \cdot 10^{-7}$ | 0.69 | -13 | -11 | $367.2 \cdot 10^{-11}$ | 2.4 | -30 | -35 |
| 325.00 | $-151.5 \cdot 10^{-7}$ | 1.2 | -17 | -15 | $42.9 \cdot 10^{-10}$ | 2.4 | -15 | -21 |
| 350.00 | $-68.3 \cdot 10^{-7}$ | 4.7 | -34 | -31 | $38.5 \cdot 10^{-10}$ | 5.1 | -20 | -27 |
| 375.00 | $-25.9 \cdot 10^{-7}$ | 6.9 | -26 | -21 | $559.2 \cdot 10^{-11}$ | 1.3 | 20 | 8.5 |
| | | | | ($0.90\ CH_4 + 0.10\ He$) | | | | |
| 273.16 | $-283.9 \cdot 10^{-7}$ | 0.92 | -13 | -8.7 | $47.8 \cdot 10^{-10}$ | 3.9 | -4.2 | -14 |
| 300.00 | $-161.0 \cdot 10^{-7}$ | 1.1 | -24 | -18 | $367.2 \cdot 10^{-11}$ | 2.6 | -22 | -30 |
| 325.00 | $-76.6 \cdot 10^{-7}$ | 2.4 | -40 | -31 | $38.7 \cdot 10^{-10}$ | 2.6 | -14 | -23 |
| 350.00 | $-12.4 \cdot 10^{-7}$ | 24 | -78 | -70 | $43.0 \cdot 10^{-10}$ | 4.0 | -0.3 | -12 |
| 375.00 | $42.2 \cdot 10^{-7}$ | 4.5 | 677 | 145 | $485.0 \cdot 10^{-11}$ | 1.6 | 17 | 1.2 |

**Table 10.** Statistical analysis of the speed of sound data with respect to AGA8-DC92 and GERG-2008 EoS for the three binary ($CH_4$ + He) mixtures of this research. $\Delta_{AAD}$ = average absolute relative deviation, $\Delta_{Bias}$ = average relative deviation, $\Delta_{RMS}$ = root mean square relative deviation, $\Delta_{MaxD}$ = maximum relative deviation.

| | $10^2 \cdot$(Experimental vs AGA) | | | | $10^2 \cdot$(Experimental vs GERG) | | | |
|---|---|---|---|---|---|---|---|---|
| | $\Delta_{AAD}$ | $\Delta_{Bias}$ | $\Delta_{RMS}$ | $\Delta_{MaxD}$ | $\Delta_{AAD}$ | $\Delta_{Bias}$ | $\Delta_{RMS}$ | $\Delta_{MaxD}$ |
| ($0.95\ CH_4 + 0.05\ He$) | 0.12 | 0.091 | 0.13 | 0.19 | 0.38 | 0.36 | 0.43 | 0.61 |
| ($0.90\ CH_4 + 0.10\ He$) | 0.17 | 0.092 | 0.20 | 0.31 | 0.66 | 0.61 | 0.76 | 1.1 |

**Figure 1.** Schematic diagram of the acoustic cavity and thermostat device of the experimental setup: 1 - spherical resonance cavity, 2 – acoustic transducers, 3 – thermometers (SPRTs), 4 – copper block, 5- internal shell, 6 – external shell, 7 – gas inlet duct, 8 – to vacuum.

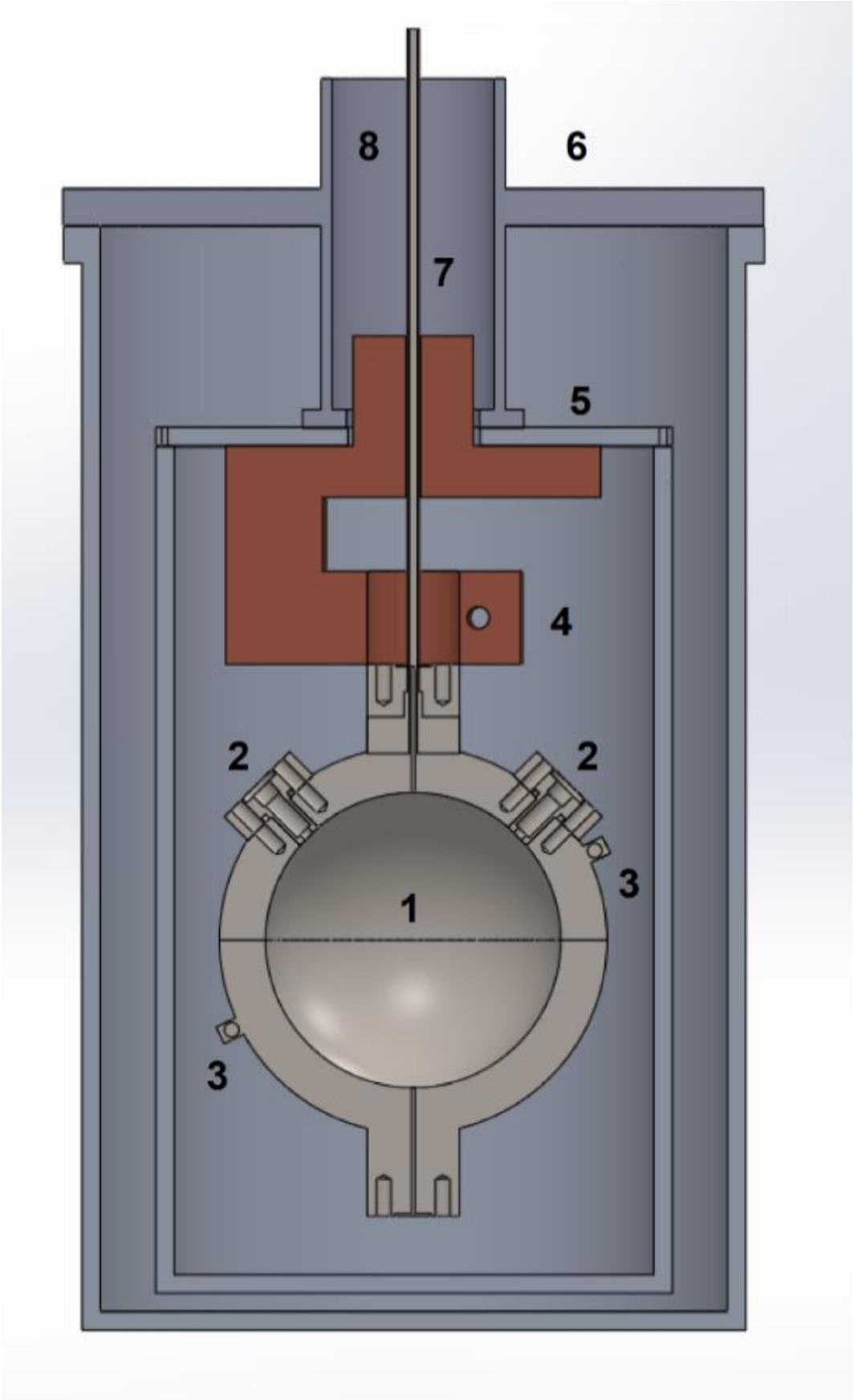

**Figure 2.** Relaxation constant times $\tau_{vib}$ due to vibrational relaxation of radial modes as a function of pressure at $T$ = 273.16 K for binary mixtures (a): (0.95 $CH_4$ + 0.05 He) and (b): (0.90 $CH_4$ + 0.10 He) and for modes: △ (0,2), ◇ (0,3), □ (0,4), × (0,5), + (0,6).

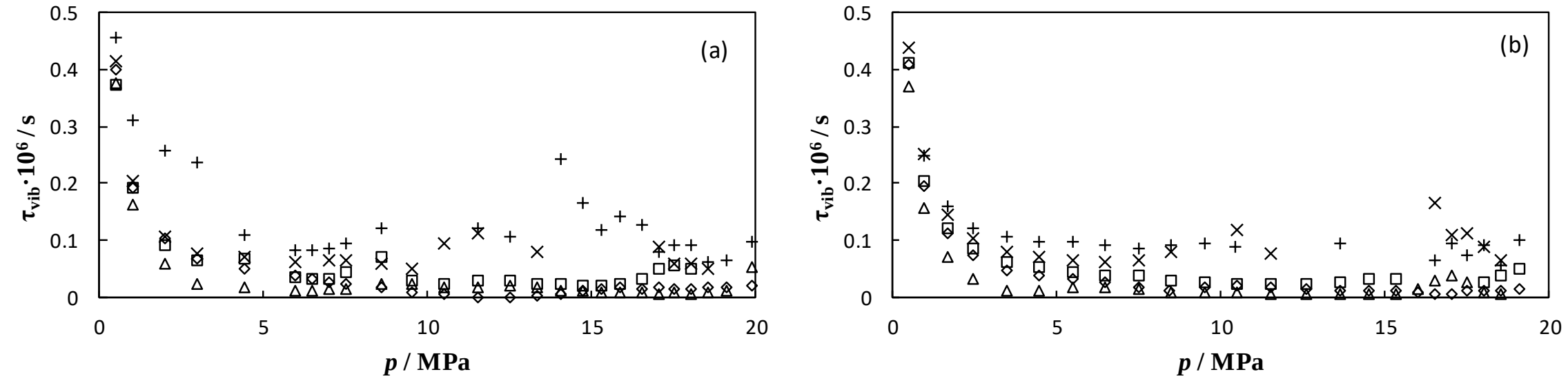

**Figure 3.** Relative excess halfwidths ($\Delta g/f$) of radial modes as a function of pressure at $T$ = 273.16 K for binary mixtures (a): (0.95 $CH_4$ + 0.05 He) and (b): (0.90 $CH_4$ + 0.10 He) and for modes: △ (0,2), ◇ (0,3), □ (0,4), × (0,5), + (0,6).

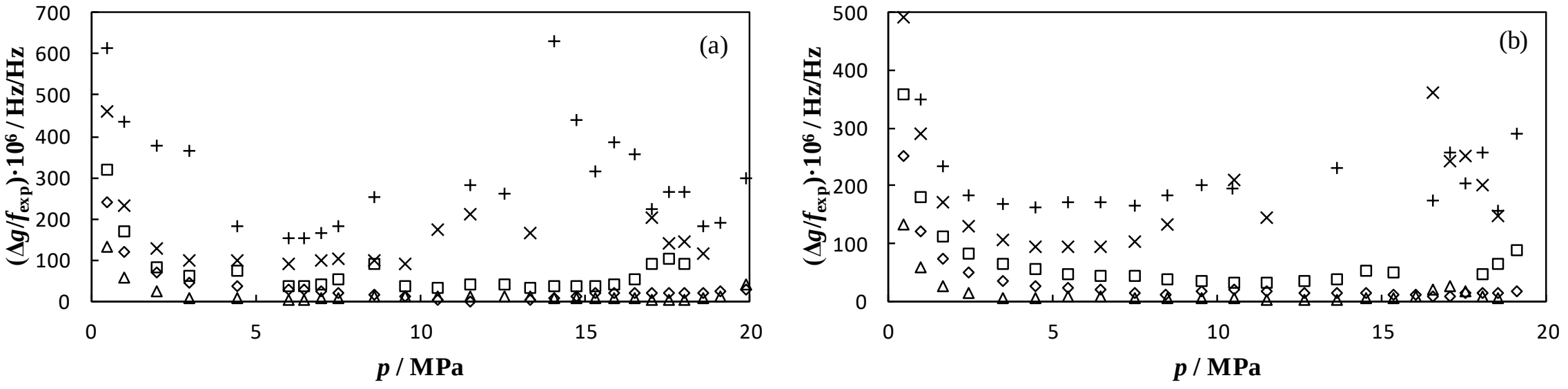

**Figure 4.** Residual analysis $\Delta w = (w_{fitted} - w_{exp})/w_{exp}$ as a function of pressure of the measured speed of sound and the values fitted by equation (25), for binary mixtures (a): (0.95 $CH_4$ + 0.05 He) and (b): (0.90 $CH_4$ + 0.10 He) at temperatures: △ 273.16 K, ◇ 300 K, □ 325 K, × 350 K, + 375 K.

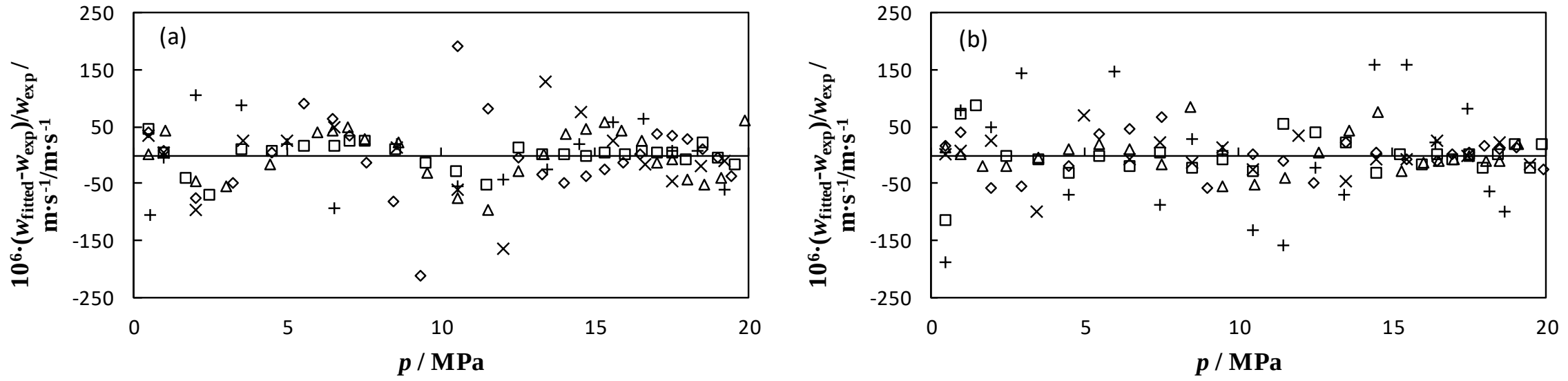

**Figure 5.** Relative deviations $\Delta w = (w_{exp} - w_{EoS})/w_{EoS}$ as function of pressure for binary mixture (0.95 $CH_4$ + 0.05 He) from calculated values from: (a): AGA8 EoS and (b): GERG-2008 EoS, expanded ($k$ = 2) experimental uncertainty in speed of sound as a dotted line and the expanded ($k$ = 2) uncertainty of model EoS as a dashed line at temperatures: △ 273.16 K, ◇ 300 K, □ 325 K, × 350 K, + 375 K.

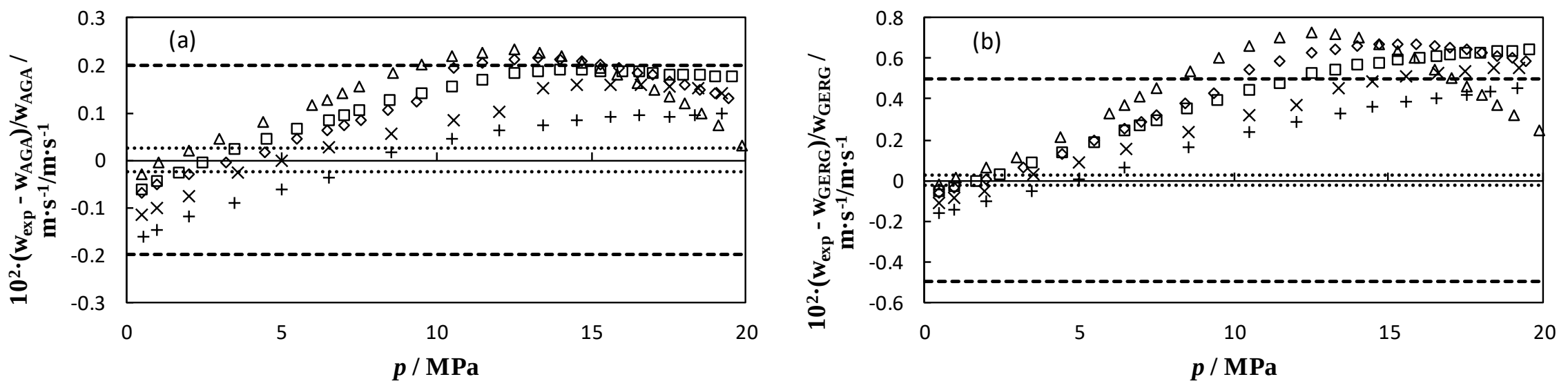

**Figure 6.** Relative deviations $\Delta w = (w_{exp} - w_{EoS})/w_{EoS}$ as function of pressure for binary mixture (0.90 $CH_4$ + 0.10 He) from calculated values from: (a): AGA8 EoS and (b): GERG-2008 EoS, expanded ($k$ = 2) experimental uncertainty in speed of sound as a dotted line and the expanded ($k$ = 2) uncertainty of model EoS as a dashed line at temperatures: △ 273.16 K, ◇ 300 K, □ 325 K, × 350 K, + 375 K.

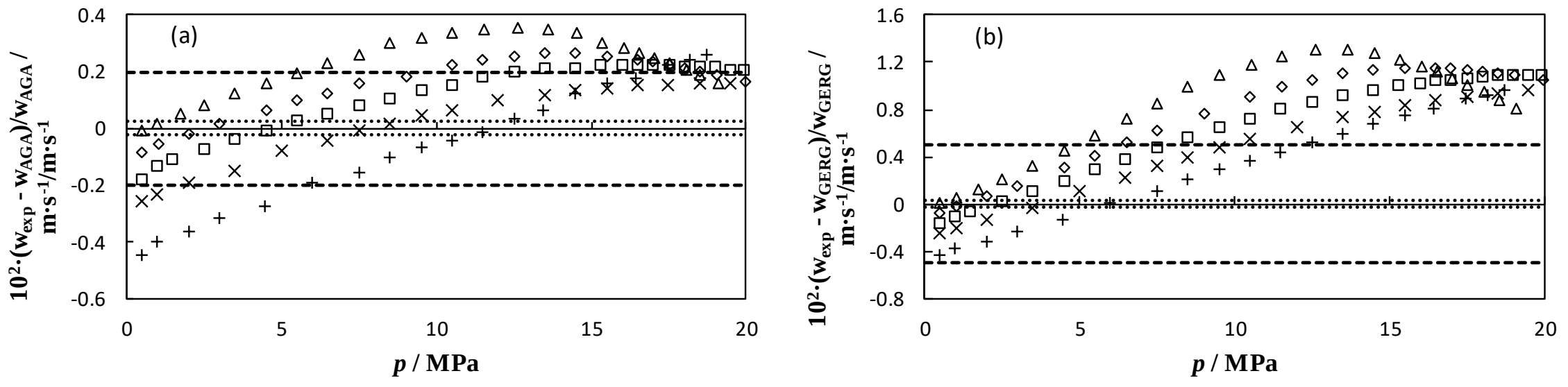

**Figure 7.** Frequency measures as function of time (in days) performed for the assessment of the stability of the mixtures studied. They correspond to the acoustic (0,3) mode at $p$ ~ 5 MPa and $T$ = 273.16 K for the binary (0.90 $CH_4$ + 0.10 He) mixture.

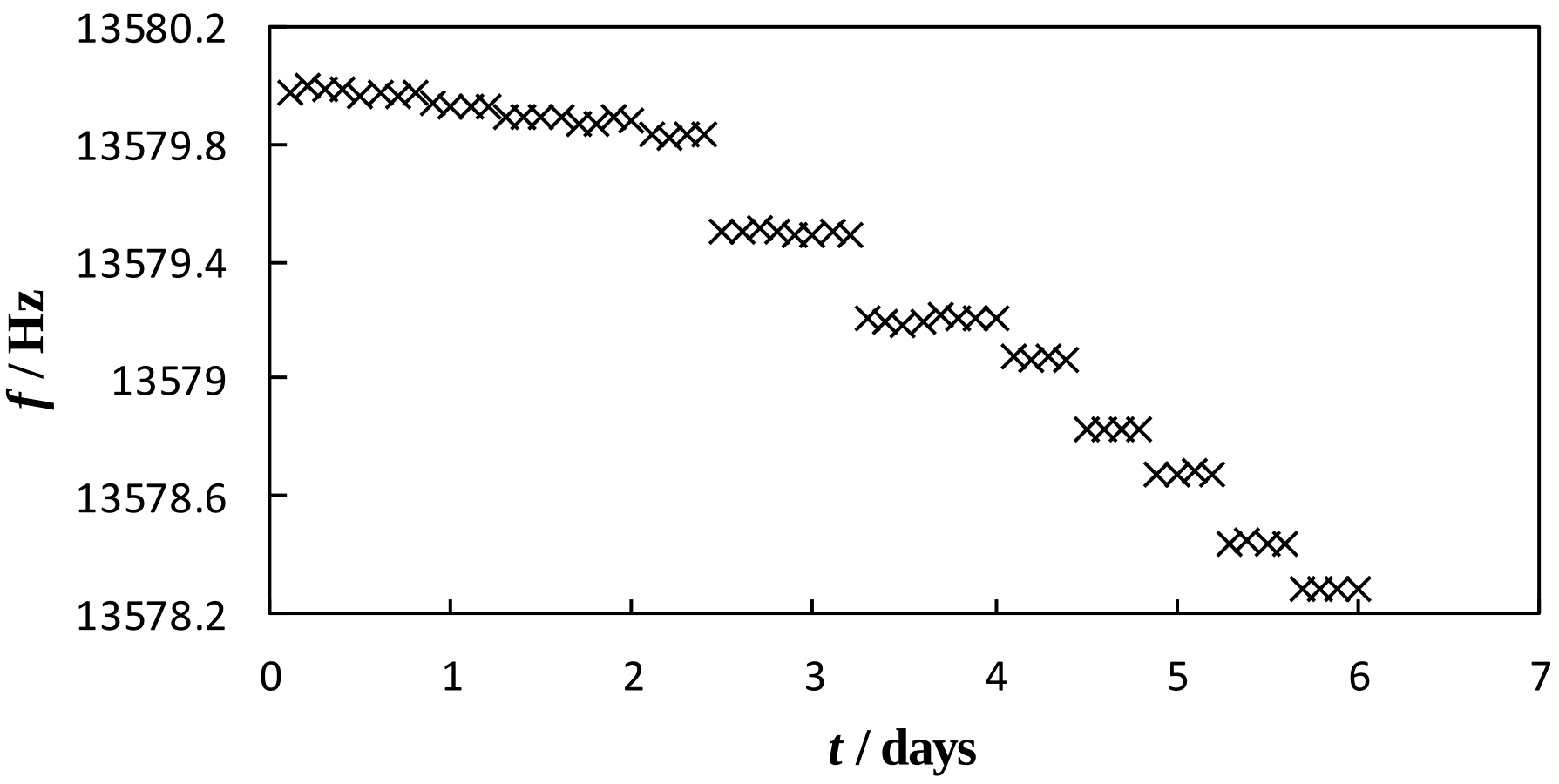